\documentclass[iop]{emulateapj}

\usepackage{graphicx}
\usepackage{pslatex}
\usepackage[T1]{fontenc}
\usepackage{type1cm}
\usepackage{color}
\usepackage{aas_macros}
\usepackage{amssymb,amsmath,amsxtra,amsfonts}
\usepackage{float}
\usepackage{natbib}
\usepackage{xspace}
\usepackage[pdf]{pstricks}

\newcommand{\fif}{4U~1538$-$522\xspace}
\newcommand{\msol}{$M_{\odot}$\xspace}
\newcommand{\rsol}{$R_{\odot}$\xspace}
\newcommand{\lsol}{$L_{\odot}$\xspace}
\newcommand{\qvnor}{QV Nor\xspace}
\newcommand{\suz}{\emph{Suzaku}\xspace}
\newcommand{\nh}{$N_{\mathrm{H}}$\xspace}
\newcommand{\alfven}{Alfv\'{e}n\xspace}

\begin{document}

\title{A Clumpy Stellar Wind and Luminosity-Dependent Cyclotron Line Revealed by The First Suzaku Observation of the High-Mass X-ray Binary 4U~1538$-$522}
\accepted{2 July 2014}

\author{Paul~B.~Hemphill \altaffilmark{1},
Richard~E.~Rothschild \altaffilmark{1},
Alex~Markowitz \altaffilmark{1,2,6},
Felix~F\"{u}rst \altaffilmark{3},
Katja~Pottschmidt \altaffilmark{4,5},
J\"{o}rn Wilms \altaffilmark{2}
}
\altaffiltext{1}{Center for Astrophysics and Space Sciences, University of California, San Diego, 9500 Gilman Dr., La Jolla, CA 920093-0424, USA}
\altaffiltext{2}{Dr.\ Karl Remeis-Sternwarte \& Erlangen Center for Astroparticle Physics, Sternwartstr. 7, 96049 Bamberg, Germany}
\altaffiltext{3}{Cahill Center for Astronomy and Astrophysics, California Institute of Technology, MC 290-17, 1200 E. California Blvd., Pasadena, CA 91125, USA}
\altaffiltext{4}{Center for Space Science and Technology, University of Maryland Baltimore County, 1000 Hilltop Circle, Baltimore, MD 21250, USA}
\altaffiltext{5}{CRESST and NASA Goddard Space Flight Center, Astrophysics Science Division, Code 661, Greenbelt, MD 20771, USA}
\altaffiltext{6}{Alexander von Humboldt fellow}

\email{pbhemphill@physics.ucsd.edu}

\begin{abstract}
  We present results from the first \suz observation of the high-mass X-ray
  binary \fif. The broad-band spectral coverage of \suz allows for a detailed
  spectral analysis, characterizing the cyclotron resonance scattering feature
  at $23.0 \pm 0.4$\, keV and the iron K$\alpha$ line at $6.426 \pm
  0.008$\,keV, as well as placing limits on the strengths of the iron K$\beta$
  line and the iron K edge. We track the evolution of the spectral parameters
  both in time and in luminosity, notably finding a significant positive correlation
  between cyclotron line energy and luminosity. A dip and spike in the
  lightcurve is shown to be associated with an order-of-magnitude increase in
  column density along the line of sight, as well as significant variation in
	the underlying continuum, implying the accretion of a overdense region of
	a clumpy stellar wind. We also present a phase-resolved analysis, with most
	spectral parameters of interest showing significant variation with phase.
	Notably, both the cyclotron line energy and the iron K$\alpha$ line intensity
	vary significantly with phase, with the iron line intensity significantly
	out-of-phase with the pulse profile. We discuss the implications of these
	findings in the context of recent work in the areas of accretion column
	physics and cyclotron resonance scattering feature formation.
\end{abstract}


\maketitle

\section{Introduction}
\label{sec:intro}

\fif is a persistent high mass X-ray binary (HMXB) discovered in the third
\textit{Uhuru} survey \citep{giacconi_third_1974}, consisting of a wind-accreting
X-ray pulsar accompanied by the B0Iab supergiant \qvnor
\citep{reynolds_optical_1538_1992}. Following the identification of the source
as a pulsar by \citet{becker_1538_1977} and \citet{davison_1538_1977b}, the
pulse period has undergone two major torque reversals, both unobserved: one in
$\sim 1990$ at a pulse period of $\sim 530.5$\,s \citep{rubin_observation_1997}
and one in $\sim 2009$ at a pulse period of $\sim 525$\,s
\citep{finger_gbm_2009}. The source is currently following a spin-down trend, as
observed by the \textit{Fermi} GBM \citep{finger_gbm_2009} and \textit{INTEGRAL}
\citep{hemphill_2013}. A $\sim 3.7$\,day orbital period was observed by
\citet{becker_1538_1977} and \citet{davison_1538_1977b}; while subsequent
analyses have not found significant changes in the orbital period, the
eccentricity of the orbit has proved to be an elusive target, with
\citet{makishima_spectra_1987} adopting $e = 0.08 \pm 0.05$ while
\citet{clark_orbit_2000} provides parameters for circular and elliptical ($e
\sim 0.17$) orbits, with the elliptical solution being updated by
\citet{mukherjee_orbital_2006}. Relatively recent work in
\citet{rawls_mass_2011} determined the mass of the compact object in \fif to be
$0.87 \pm 0.07$\,\msol for \citeauthor{clark_orbit_2000}'s elliptical orbit, and
$1.104 \pm 0.177$ for the circular solution - both substantially lower than the
canonical neutron star mass of 1.4\,\msol. Despite this very low mass, the
source is clearly a neutron star, as its magnetic field, as determined from the
energy of its cyclotron resonance scattering features, is $\sim 2 \times
10^{12}$\,G \citep{clark_discovery_1990}, significantly higher than the $\sim
10^{6}-10^{9}$\,G fields seen in white dwarfs \citep{schmidt_wd_2003}. \fif's
distance is somewhat uncertain, with \citet{crampton_1538_1978} estimating it to
be $5.5 \pm 1.5$\,kpc, \citet{ilovaisky_1538_1979} finding $6.0 \pm 0.5$\,kpc,
and \citet{reynolds_optical_1538_1992} finding $6.4 \pm 1.0$\,kpc. We adopt the
latter distance in this work.

The neutron star is a persistent X-ray emitter with occasional flaring, due to
the close and low-eccentricity nature of the binary system. Due to the lack of
a physical model, the spectrum of \fif is modeled similarly to other HMXBs, with
one or more \citep{rodes-roca_detecting_2010} absorbed power laws modified by
a high-energy exponential cutoff. An absorption feature at $\sim 22$\,keV in the
spectrum of \fif was discovered and identified as a cyclotron resonance
scattering feature (CRSF) by \citet{clark_discovery_1990}; another absorption
feature at $\sim 50$\,keV was observed by \citet{robba_bepposax_2001} and
identified as the harmonic of the 22\,keV CRSF by \citet{rodes-roca_first_2009}.
The spectrum also features a $6.4$\,keV emission feature, identified as the Fe
K$\alpha$ line \citep{makishima_spectra_1987}, as well as other emission lines
in the $\sim 1-3$\,keV and $\sim 6-7$\,keV range
\citep{rodes-roca_detecting_2010}. The parameters of the cyclotron line have
until now not displayed any detectable correlation with luminosity; however,
conclusions in this area have been difficult to arrive at due to previous work
using a mix of different empirical models for the continuum and the cyclotron
line shape, with \citet{clark_discovery_1990}, \citet{mihara_cyclovar_1998}, and
\citet{rodes-roca_first_2009} using Lorentzian-shaped line profiles for the CRSF
and \citet{coburn_magnetic_2001}, \citet{robba_bepposax_2001}, and
\citet{hemphill_2013} adopting Gaussian profiles. This is a common issue in the
study of HMXBs in general, and as shown by \citet{muller_nocorrelation_2012} in
the case of 4U~0115+63, the choice of model (both for the continuum and the
CRSF) can have large effects on the observed trend with luminosity. The factor
of $\sim 4$ luminosity range covered in this single observation allows us to
obtain a more self-consistent picture of the behavior of the cyclotron line.

\section{Observation}
\label{sec:observation}

\suz observed \fif on 10 August 2012 starting at UTC 00:04 (MJD 56149.003) for
$61.9$\,ks. The observation ID was 407068010.  \suz carries two sets of
operational instruments: the X-ray Imaging Spectrometer
\citep[XIS,][]{xis_2007}, consisting of four CCD detectors (XIS0-3) covering
a $0.2-12$\,keV bandwidth, and the Hard X-ray Detector \citep[HXD,][]{hxd_2007},
which is comprised of a set of silicon PIN diodes (energy range $10-70$\,keV)
and the GSO scintillator ($40 - 600$\,keV).  The observation was carried out in
the XIS-nominal pointing mode. One XIS unit, XIS2, is no longer operational,
being taken offline after a micrometeorite impact in November 2006, so only the
front-illuminated XIS0 and XIS3 and the back-illuminated XIS1 were used in the
observation. We additionally chose not to use data from the GSO scintillators
due to very low signal compared to background in that instrument. XIS and
HXD/PIN data were reprocessed starting from the ``unfiltered'' event files using
the standard \suz tools provided in the HEASOFT software suite, version 6.13.
The extraction software makes use of the calibration database (CALDB) maintained
by HEASARC; our extraction used version 20130916 for the XIS and 20110913 for
the HXD/PIN.

\subsection{XIS data reduction}
\label{ssec:xis_red}

To mitigate pile-up in the detectors, the three XIS instruments operated in 1/4
window mode. Each XIS unit features $^{55}$Fe calibration sources illuminating
two corners of each CCD, but this windowing mode precluded their use. An
additional attitude correction was performed on the event files using the \suz
tool \texttt{aeattcor2} prior to spectral extraction. A pileup estimation using
the \texttt{pileest} tool showed maximum pileup of $\sim 7$\% around the center
of the image, and so the source regions for each XIS unit were defined as
annuli, excising the central region down to $\sim 3$\% pileup. Background
regions were defined as strips along the edges of each image; in the case of
XIS0, the background regions additionally avoided a strip of bad pixels located
along one edge of the image \citep[due to what was likely a micrometeorite
impact in 2009 - see][]{jx_isas_suzaku_memo_2010-01}. In all detectors,
background regions were defined with the same total area as the source region.
Source and background spectra and lightcurves for each XIS unit were extracted
using \texttt{xselect}, totaling 46\,ks of exposure in each detector, and
response matrices (RMFs) and ancillary response files (ARFs) were generated
using \texttt{xisrmfgen} and \texttt{xissimarfgen}. Source spectra were
regrouped using \texttt{grppha}; the grouping used is the same as that used by
\citet{nowak_xisbinning_2011}, which attempts to optimally account for the
energy resolution of the XIS CCDs. Data below 1\,keV and between 1.6 and
2.3\,keV showed significant discrepancies between XIS1 and XIS0+XIS3, and were
ignored; we additionally ignored XIS data above 10\,keV.

\subsection{PIN data reduction}
\label{ssec:pin_red}

Data from the \suz HXD/PIN were extracted using \texttt{xselect}. The ``tuned''
non-X-ray background (NXB) event files provided by the \suz team were used,
while the cosmic X-ray background (CXB) was simulated according to the model of
\citet{boldt_cxb_1987}, using the \texttt{fakeit} procedure in XSPEC v.12.8.0.
The NXB and CXB spectra were added using \texttt{mathpha} to produce a single
PIN background spectrum. The source spectrum was deadtime-corrected using the
\texttt{hxddtcor} tool; the total PIN exposure after deadtime correction is
34\,ks. PIN spectra were regrouped to 100 counts/bin, and data were ignored
below 15\,keV in all datasets and above $\sim 40 - 50$\,keV, depending on the
signal-to-noise in the particular dataset being analyzed.

\section{Timing Analysis}
\label{sec:lc}
\begin{figure}[ht!]
	\centering
	\epsscale{0.85}
	\plotone{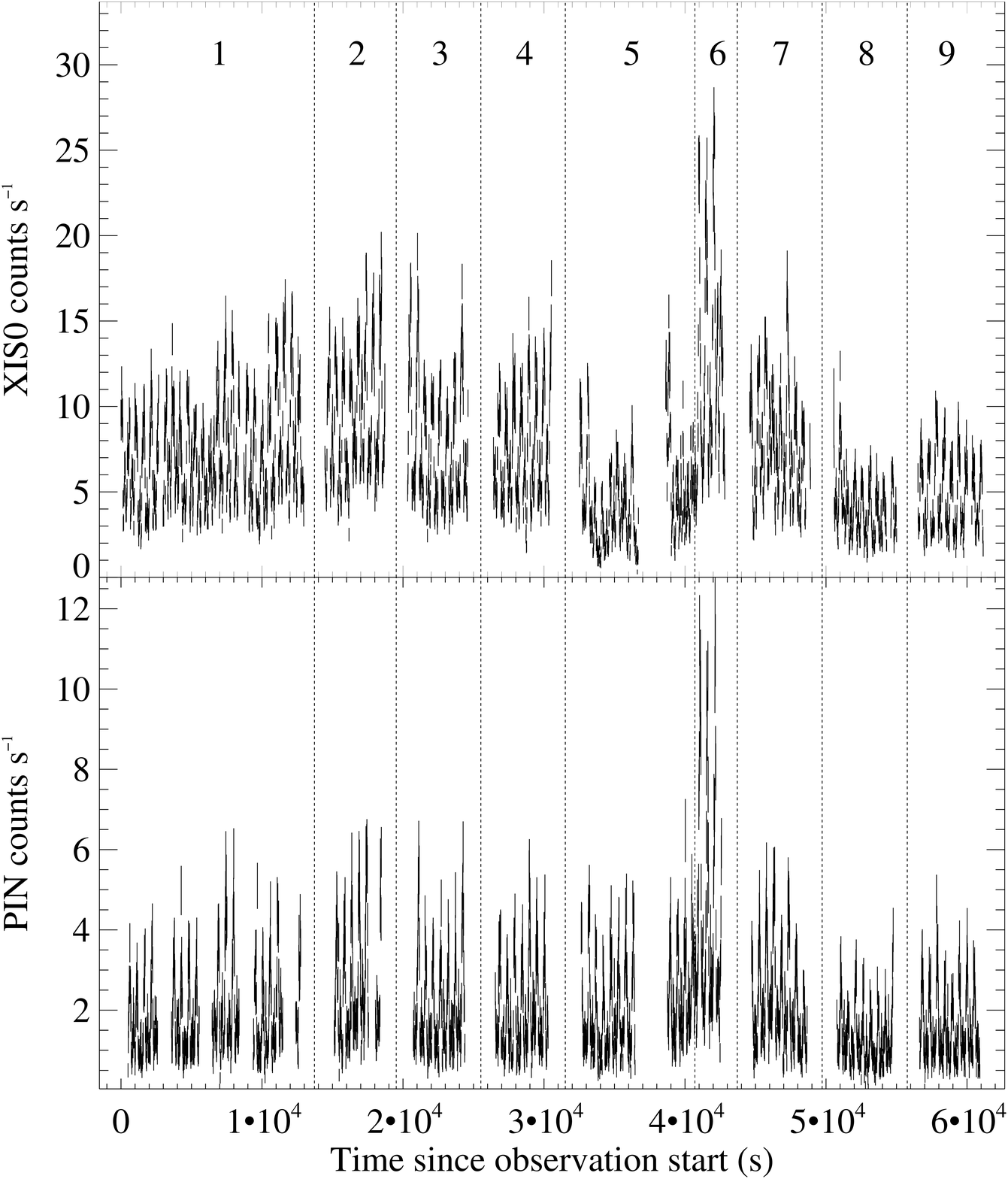}
	\caption{The XIS0 (top) and HXD/PIN (bottom) lightcurves for the observation.
		The time resolution for both lightcurves is 16\,s, and both use the full
		energy range of the detectors (0.1 to 12\,keV for XIS0, 10 to 70\,keV for
		the HXD/PIN). The large gaps every $\sim 6$\,ks are due to the satellite's
		passage through the South Atlantic Anomaly. The vertical lines and numbers
		indicate the time bins for the time-resolved analysis; the time-averaged
		spectra use data from bins 1 through 4.}
  \label{fig:lightcurve}
\end{figure}

We extracted 2\,s-binned lightcurves from the XIS units and 16\,s-binned
lightcurves from both the XIS and the PIN. The 16\,s-binned XIS0 and PIN
lightcurves are plotted in Figure~\ref{fig:lightcurve}. Aside from the regular
pulsations from the rotation of the neutron star, which are easily seen
throughout the observation, there is additional clear structure in the
lightcurve.  A significant dip in flux is visible in the XIS lightcurve at $\sim
3.5 \times 10^{4}$\,s into the observation; this feature is not visible in the
PIN lightcurve. A flare at $\sim 4 \times 10^4$\,s is visible in both
instruments, and represents an increase in flux by a factor of $\sim 3$ for
approximately three pulses. The counting rate drops during the post-flare
portion of the observation, but unlike the dip, this decrease appears in both
the XIS and the PIN and likely reflects an overall decrease in the accretion
rate. The origins of these features can be better seen with a spectral analysis
(see Section~\ref{sec:spec}). The pulse period is $525.59 \pm 0.04$\,s,
determined via epoch folding \citep{leahy_epfold_1983,larsson_epfold_1996} using
the 2\,s-binned XIS0 lightcurve.  \textit{Fermi} GBM monitoring \citep[as
originally described in ][]{finger_gbm_2009} of the source finds a similar
period of $\sim 525.7$\,s at around the time of the \textit{Suzaku}
observation\footnote{For the most recent results, see
\url{http://gammaray.nsstc.nasa.gov/gbm/science/pulsars/}}.

\begin{figure}[ht!]
	\centering
	\epsscale{1.0}
	\plotone{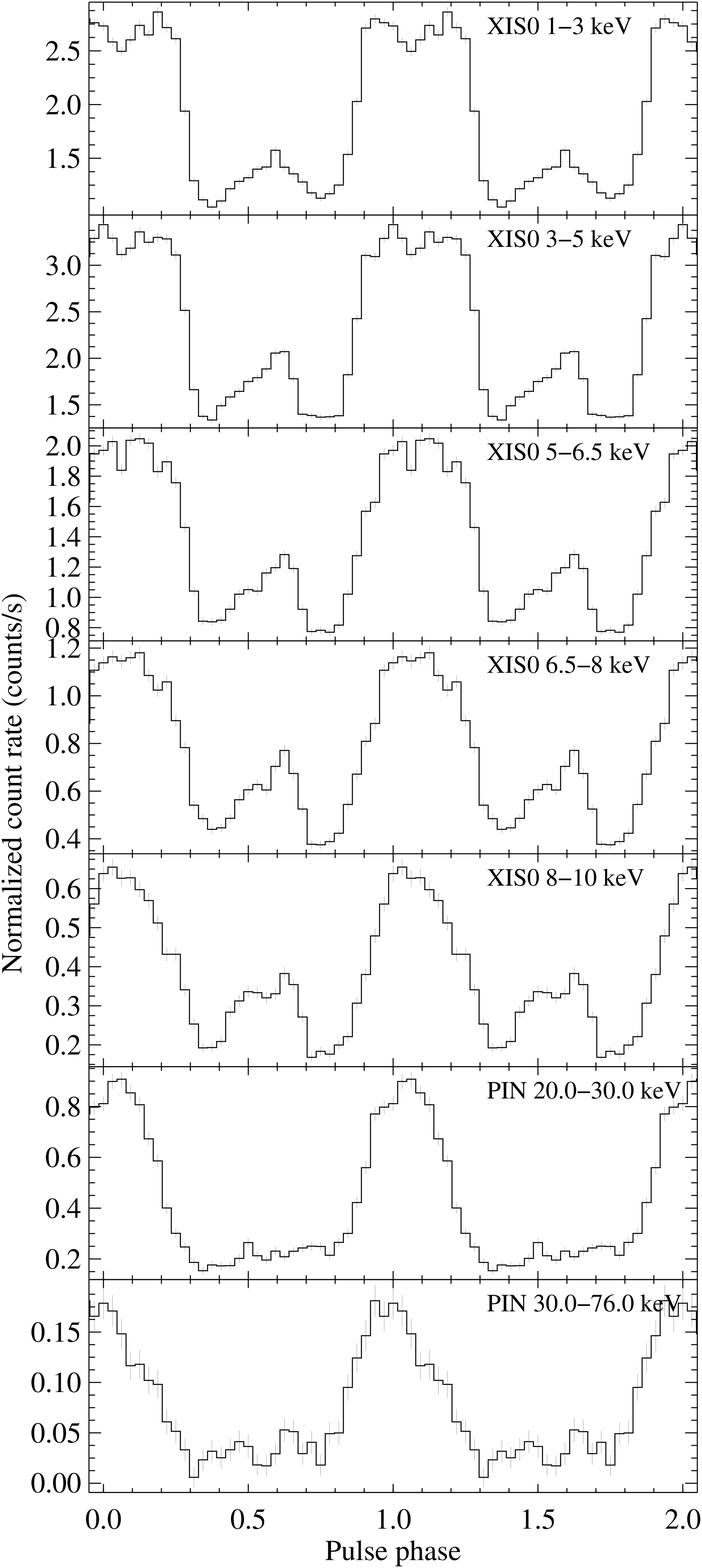}
	\caption{Energy-resolved pulse profiles for \fif. The source lightcurves have
		been shifted so that phase 0 represents the same time for each profile.
		There is evidence of a phase shift in the peak of the main pulse, which can
		be seen as the main pulse consisting of two separate peaks blended together.
		The weak secondary peak disappears above $\sim 20$\,keV, which may be in part
		due to the presence of the CRSF at $\sim 23$\,keV.}
	\label{fig:prof}
\end{figure}

We extracted lightcurves in multiple energy bands and folded them on the
determined pulse period to obtain the pulse profile in each energy band. These
profiles can be found in Figure~\ref{fig:prof}. The pulse profile is
double-peaked at lower energies, with a weak secondary peak that essentially
disappears at energies above $\sim 20$\,keV.

The main peak is flat-topped at low energies, narrowing and becoming more peaked
as energy increases. These changes are, however, quite slight in comparison to
other sources \citep[as seen in, e.g., 4U~0115+63 by][]{ferrigno_0115_2011}. The
main peak additionally appears to shift slightly in phase, with the main pulse
of the high-energy pulse profiles slightly preceding the center of the main
pulse in the lower-energy profiles. While energy-dependent lags are not unknown
in X-ray pulsars, especially at around the cyclotron line energy
\citep{ferrigno_0115_2011}, such phase lags have not been observed before in
\fif \citep{clark_discovery_1990,coburn_magnetic_2001,robba_bepposax_2001}. The
behavior of the main pulse can be modeled with two blended peaks, one relatively
energy-independent peak at phase $\sim 0.9$ and one at phase $\sim 0.15$ which
decreases in height with increasing energy. The ``knee'' at phase $\sim 0.1$
visible in the 30--76\,keV profile may be indicative of more complex energy
dependence in the main pulse. The overall effect is an asymmetry in spectral
hardness across the main pulse, which is easily seen in the phase-resolved
spectra discussed in Section~\ref{ssec:phase}.

\section{Spectral Analysis}
\label{sec:spec}
We analyze the spectrum of \fif in four different ways. First, we examine the
time- and phase-averaged spectrum from the first 32\,ks of the observation
(Section~\ref{ssec:fullspec}). Second, we divided the observation into pieces
with exposures of $\sim 4$\,ks, in order to perform a coarsely-binned
time-resolved analysis of the broad-band spectrum (Section~\ref{ssec:time}).
A more finely-binned analysis of the XIS spectra of the individual pulses
(Section~\ref{ssec:pulse}) gives a much higher resolution picture of the
behavior of the lower-energy spectrum of the source.  Finally, via
phase-resolved analysis (Section~\ref{ssec:phase}), we investigate the source's
behavior over its pulse profile. The time intervals for the time-averaged and
time-resolved spectra are indicated in Figure~\ref{fig:lightcurve}. In the full
time-averaged spectrum, the coarsely-binned time-resolved spectra, and the
phase-resolved spectra, all three XIS units and the HXD/PIN were fit
simultaneously in XSPEC version 12.8.0. In the pulse-to-pulse analysis, only XIS
spectra were used, as the low exposure for each individual spectrum precluded
the use of PIN data.

\subsection{Time-averaged spectrum}
\label{ssec:fullspec}
We do not examine the full observation in a single spectrum due to the high
variability seen with the onset of the dip. Rather, we investigate here the
time-averaged spectrum of the first 32\,ks of the observation, where the
source's spectral variability is relatively low (see Sections \ref{ssec:time}
and \ref{ssec:pulse}). The three XIS and single PIN spectra cover an energy
range of 1 to 40\,keV, with gaps between 1.6 and 2.3\,keV (due to uncertainties
in the instrumental response) and between 10 and 15\,keV (where both the XIS and
the PIN lack good coverage). We fit our broad-band spectra with three different
continuum models: a single power-law multiplied by either the \texttt{highecut}
\citep{white_highecut} or \texttt{fdcut} \citep{tanaka_fdcut} exponential
cutoff, and the two-power-law \texttt{npex} \citep{mihara_thesis_1995}:
\begin{eqnarray}
  \label{eqn:highecut}
  \mathtt{pow*highecut}(E) &=& \left\{\begin{array}{cc}
    A E^{-\Gamma} & E < E_{\mathrm{cut}} \\
    A E^{-\Gamma}\exp\left( \frac{E_{\mathrm{cut}} - E}{E_{\mathrm{fold}}} \right) & E \ge E_{\mathrm{cut}} \\
	\end{array}\right. \\[1ex]
  \label{eqn:fdcut}
	\mathtt{pow*fdcut}(E) &=& A E^{-\Gamma} \left[1 + \exp\left(\frac{E - E_{\mathrm{cut}}}{E_{\mathrm{fold}}}\right)\right]^{-1} \\[1ex]
  \label{eqn:npex}
  \mathtt{npex}(E) &=& A \left( E^{-\alpha} + BE^{+\beta} \right) e^{E/kT}
\end{eqnarray}
The remaining widely-used power-law/exponential cutoff continuum model,
\texttt{cutoffpl}, did not provide good fits, with significant additional
structure at higher energies and a reduced $\chi^{2}$ of $\sim 2$, and so it was
not used. In equations \ref{eqn:highecut} and \ref{eqn:fdcut}, the parameter
$\Gamma$ is the power-law index. We should
note here that, despite using the same parameter names, the $E_{\mathrm{cut}}$
and $E_{\mathrm{fold}}$ parameters of the \texttt{highecut} and \texttt{fdcut}
models are mathematically different and should not, in principle, be compared
directly (although they do play similar roles). \texttt{npex}
(Equation~\ref{eqn:npex}) has two power-laws, with the $\alpha$ index being
negative-definite and the $\beta$ index being positive-definite. The $B$
parameter is the relative normalization of the positive-index power-law. The
considerable added leverage of the second power-law tends to make fitting with
\texttt{npex} a difficult proposition; we avoid this somewhat by freezing
$\beta$ to $2.0$; this results in the model simulating the Wien hump at $E \sim
kT$ that should exist in the Comptonized spectrum of the neutron star
\citep{makishima_1907_1999}. \texttt{npex} features a $kT$ parameter, which
functions in a similar capacity to $E_{\mathrm{fold}}$ in the other two models.

\begin{figure}[ht]
	\centering
	\epsscale{1.0}
	\plotone{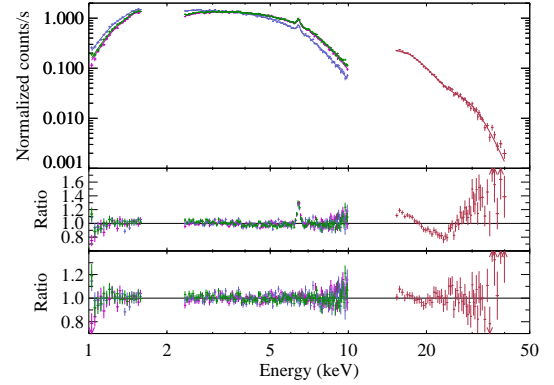}
	\caption{The time-averaged spectrum for \fif before the dip. The continuum
		model is \texttt{fdcut}. The gap between 1.6 and 2.3\,keV is due to
		uncertainties in the XIS instrumental response in that range. From top to
		bottom, we plot the spectrum with the best-fitting model, the ratio
		residuals for a simple absorbed power-law with an \texttt{fdcut} cutoff, and
		the residuals for the best-fit model, consisting of the absorbed
		power-law/\texttt{fdcut} continuum with iron K$\alpha$ and K$\beta$ lines
		and a \texttt{gauabs} feature modeling the CRSF at ${\sim 22}$\,keV.}
	\label{fig:full}
\end{figure}

Each continuum model is modified by absorption, using the latest version
\footnote{\url{http://pulsar.sternwarte.uni-erlangen.de/wilms/research/tbabs/}}
of the \texttt{tbnew} absorption model \citep{wilms_tbnew_2010}
using the abundances of \citet{wilms_tbabs} and cross sections from
\citet{verner_xsect}. In addition to the broad effects of absorption, the
residuals have significant additional structure between $6.4$ and $\sim
7.5$\,keV and at $\sim 23$\,keV. The strong emission line at $\sim 6.4$\,keV is
the K$\alpha$ emission of neutral iron, while additional emission and
absorption-like features in the residuals at $\sim 7$\,keV are identified as the
iron K$\beta$ emission line ($7.06$\,keV) and the iron K-shell ionization edge
($7.112$\,keV in neutral iron). The emission lines are modeled with additive
gaussians, while the edge feature is modeled by allowing the abundance of iron
in the absorber to vary. We fix the widths of the emission lines to $0.01$\,keV,
as the lines widths could not be constrained. We could not resolve the 6.4\,keV
feature into multiple emission lines as was done by
\citet{rodes-roca_detecting_2010}. At $\sim 23$\,keV there is a broad
absorption-like feature, which we identify as \fif's CRSF. This feature is
included in the model using a local XSPEC model, \texttt{gauabs},
a multiplicative absorption feature with a Gaussian optical depth profile:
\begin{eqnarray}
  \mathtt{gauabs}(\tau) &=& e^{-\tau\left(E\right)} \\
  \tau(E) &=& \tau_{0}\exp\left( -\frac{\left(E-E_{0}\right)^{2}}{2\sigma^{2}}\right)
  \label{eqn:gauabs}
\end{eqnarray}
Here, $E_{0}$ is the centroid energy of the feature, $\tau_{0}$ is the maximum
optical depth, and $\sigma$ is the width of the optical depth profile. The
\texttt{gauabs} component modifies only the power-law and exponential cutoff
continuum, as it is produced via the same physical processes that produce the
rest of the continuum deep in the accretion column. The two Gaussian emission
lines, Fe K$\alpha$ and K$\beta$, are added to this continuum, and the result is
then modified by the \texttt{tbnew} absorption model. Unabsorbed fluxes are
determined using the \texttt{cflux} model in XSPEC.  The full model additionally
has a multiplicative constant in order to account for calibration differences
between instruments. The normalization constant for XIS0 was frozen at 1, while
the constant for XIS1 and XIS3 were allowed to vary.  The PIN normalization was
frozen at the nominal value of $1.16$ per the Suzaku ABC guide, as the lack of
overlap between the PIN and XIS made this parameter poorly constrained (and
influential over values of the high energy cutoff parameters). The fitted
parameters with 90\% error bars can be found in Table~\ref{tab:full}, and the
best-fit spectrum with the \texttt{fdcut} continuum is plotted in
Figure~\ref{fig:full}.

\begin{deluxetable*}{llrrr}
  \tablecolumns{5}
  \tablewidth{0pt}
  \tablecaption{Spectral fits for time-averaged spectrum}
	\tablehead{ \multicolumn{2}{l}{Continuum model} & 
              \colhead{\texttt{highecut}} &
              \colhead{\texttt{fdcut}} &
              \colhead{\texttt{npex}} }
 \startdata
 Flux\tablenotemark{a}										&	$\times10^{-10}$\,erg\,cm$^{-2}$\,s$^{-1}$& $4.2 \pm 0.2$					 & $4.2 \pm 0.2$						&	$4.2 \pm 0.2$  \\
  \nh                                     & $\times 10^{22}$\,cm$^{-2}$               & $2.1 \pm  0.2$         & $2.1 \pm  0.2$           & $1.7 \pm  0.3$  \\
  $\Gamma$                                &                                           & $1.17 \pm  0.01$       & $1.16^{+ 0.01}_{- 0.02}$ & \nodata  \\
	$A$\tablenotemark{b}										& $\times 10^{-2}$\,ph\,cm$^{-2}$\,s$^{-1}$ &	$4.75 \pm 0.08$				 & $4.70 \pm 0.08$				  & $4.76 \pm 0.18$  \\
  $E_{\mathrm{cut}}$                      & keV                                       & $21^{+2}_{-3}$         & $27 \pm 1$               & \nodata  \\
  $E_{\mathrm{fold}}$                     & keV                                       & $8^{+2}_{-1}$          & $5.0 \pm  0.7$           & \nodata  \\
  $\alpha$                                &                                           & \nodata                & \nodata                  & $0.67 \pm  0.03$  \\
  $B$                                     & $\times 10^{-3}$                          & \nodata                & \nodata                  & $3.2 \pm  0.2$  \\
	$k_{\rm B}T$                                    & keV                                       & \nodata                & \nodata                  & $4.83^{+ 0.09}_{- 0.08}$  \\[1ex]
  \hline \\
  $E_{\mathrm{K}\alpha}$                  & keV                                       & $6.426 \pm  0.008$     & $6.426 \pm  0.008$       & $6.425 \pm  0.008$  \\
  $I_{\mathrm{K}\alpha}$\tablenotemark{c} & $\times 10^{-4}$\,ph\,cm$^{-2}$\,s$^{-1}$ & $3.7 \pm  0.3$         & $3.6 \pm  0.3$           & $3.7 \pm  0.3$  \\
  $E_{\mathrm{K}\beta}$                   & keV                                       & $7.05^{+0.08}_{-0.07}$ & $7.05 \pm 0.08$          & $7.06^{+0.13}_{-0.07}$ \\
  $I_{\mathrm{K}\beta}$\tablenotemark{d}  & $\times 10^{-5}$\,ph\,cm$^{-2}$\,s$^{-1}$ & $4 \pm 2$              & $3 \pm 2$                & $4 \pm 2$  \\
  Fe abundance \tablenotemark{e}          &                                           & $1.4^{+ 0.9}_{- 0.7}$  & $1.2^{+ 0.9}_{- 0.7}$    & $>1.06$  \\[1ex]
  \hline \\
  $E_{\mathrm{cyc}}$                      & keV                                       & $22.2^{+ 0.8}_{- 0.7}$ & $23.0 \pm  0.4$          & $22.4 \pm  0.3$  \\
  $\sigma_{\mathrm{cyc}}$                 & keV                                       & $3.0 \pm  0.3$         & $3.2 \pm  0.4$           & $2.9 \pm  0.4$  \\
  $\tau_{\mathrm{cyc}}$                   &                                           & $0.9^{+ 0.1}_{- 0.2}$  & $0.72^{+ 0.09}_{- 0.08}$ & $0.58 \pm  0.05$ \\[1ex]
  \hline \\[-1ex]
  $\chi^{2}_{\mathrm{red}}$ (DOF)         &                                           & $1.17 (444)$           & $1.21 (444)$             & $1.23 (444)$
 \enddata
 \tablenotetext{a}{2--10\,keV flux}
 \tablenotetext{b}{Power-law normalization at 1\,keV}
 \tablenotetext{c}{Iron K$\alpha$ intensity}
 \tablenotetext{d}{Iron K$\beta$ intensity}
 \tablenotetext{e}{Relative to ISM abundances}
 \label{tab:full}
\end{deluxetable*}

\subsection{Time-resolved spectroscopy}
\label{ssec:time}
Gaps in the lightcurve, caused by the passage of the satellite through the
South Atlantic Anomaly, provide convenient points for a time-resolved spectral
analysis. We thus divide the observation into nine segments, defined by the SAA
gaps. The exception to this is for the fifth and sixth time bins, where the
dividing line is placed at the start of the flare rather than during the SAA
gap. This ensures that the fifth bin only encompasses the dip, and the sixth,
the flare. The time intervals for these spectra are indicated in
Figure~\ref{fig:lightcurve}. The HXD/PIN and three XIS spectra for each time bin
were initially fit with the same set of models as used for the time-averaged
spectrum. The iron K$\beta$ was generally not significantly detected in these
spectra, but we did include it in the model (with its energy fixed at
$7.056$\,keV) in order to obtain upper limits on its intensity.  The iron
abundance was additionally fixed to the ISM abundance, as it was unconstrained at
the high end.

All three continuum models fit the pre-dip and spike spectra well, with
$\chi^{2}_{\rm red} \sim 1.0 - 1.1$. The dip and the final two spectra (spectra 4,
8, and 9\ in Table \ref{tab:time}), however, show noticeable curvature in their
XIS spectra and have much poorer fits, with the dip spectrum having a reduced
$\chi^{2}$ of 1.6 for 427 degrees of freedom and spectra 8 and 9 having reduced
$\chi^{2}$ of $\sim 1.3$ for 415 and 421 degrees of freedom, respectively. We
thus modify the model for these spectra by applying a partial-covering
absorption component of the form
\begin{equation}
	\mathtt{partcov}(E) = (1-f_{\text{pcf}})*\mathtt{tbnew}_{1}(E) + f_{\text{pcf}}*\mathtt{tbnew}_{2}(E)
  \label{eqn:partcov}
\end{equation}

This approximates the effect of the source being obscured by a varying absorbing
column density. The entire source is absorbed by some base column density \nh
(this corresponds to the value of \nh measured by the single \texttt{tbnew}
component used in the other spectra), while a region with a higher column
density of $N_{\rm H,pcf}$ covers some fraction $f_{\rm pcf}$ of the light from
the source. This modification of the model is quite successful, bringing the
value of reduced $\chi^{2}$ for the dip spectrum and spectrum 9 down to
$\sim 1$ for each continuum model. The lowest-luminosity spectrum, number 8,
still has the highest overall $\chi^{2}$ even with the partial-covering model
applied, with $\chi^{2}_{\rm red} = 1.20$ for 413 degrees of freedom. We
attempted to apply the partial-covering model to the remaining spectra, but the
covering fraction could not be constrained and the quality of the fit was not
improved. In the final two spectra, the addition of the partial-covering model
increases the power-law index significantly relative to its value when only
a single absorber is used, and the high-energy cutoff parameters are found at
values that are more in line with their pre-dip values. While this means
comparing the parameters of these spectra to those without the partial-covering
model is somewhat risky, it should be noted that our pulse-to-pulse analysis
(Section \ref{ssec:pulse}) finds broadly similar results in terms of \nh and the
power-law index.

\begin{figure*}[ht]
  \centering
	\epsscale{0.9}
  \plottwo{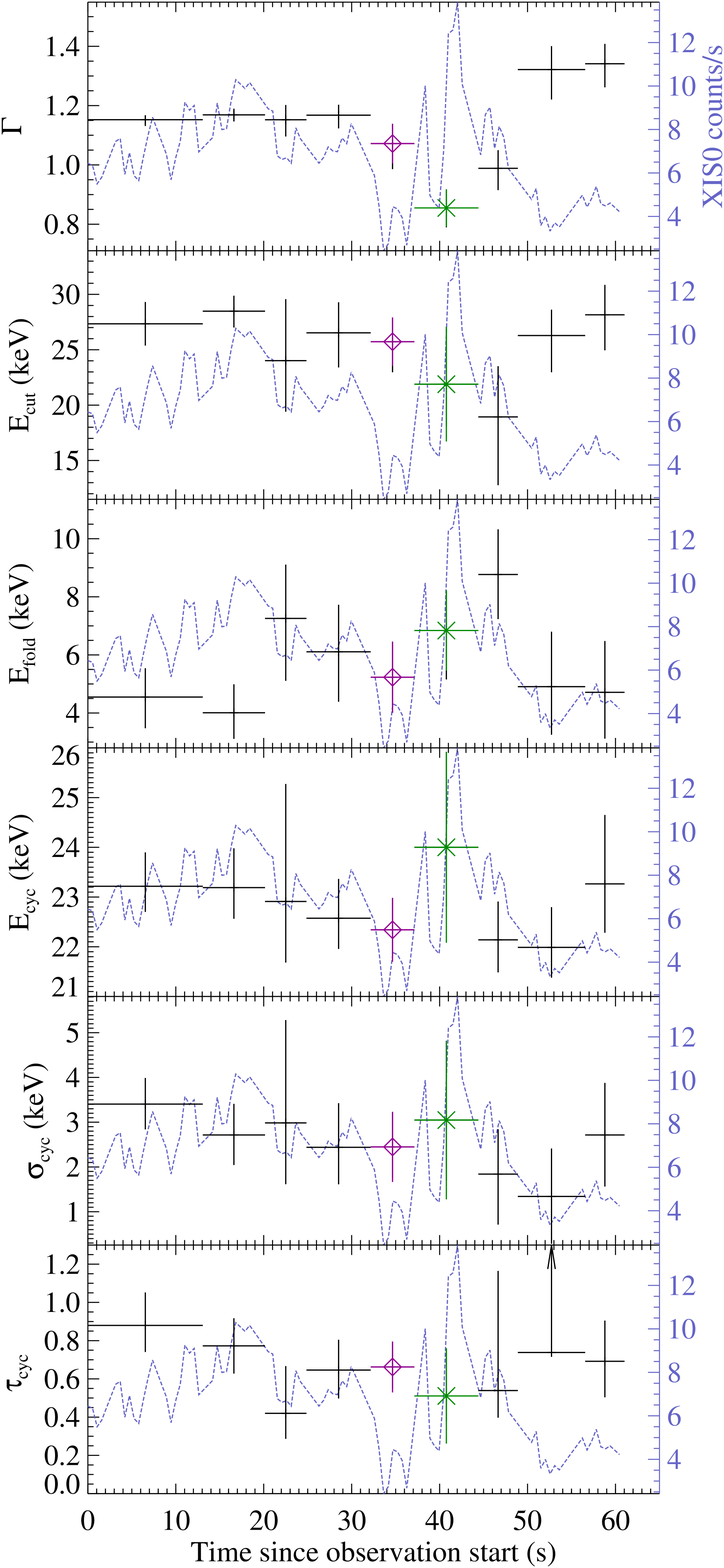}{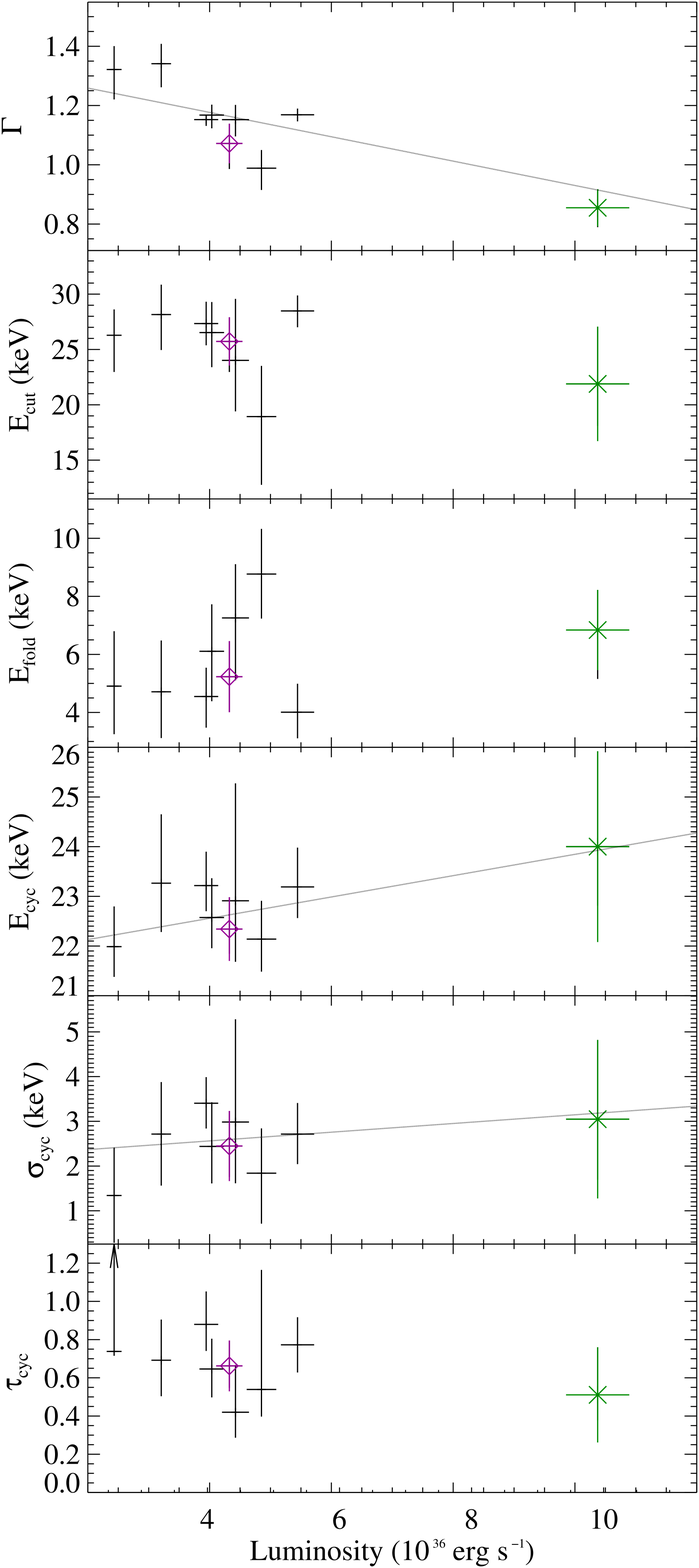}
	\caption{Selected spectral parameters over the observation, plotted against
		time (left) and against luminosity (right). For clarity, we only plot
		results from the \texttt{fdcut} continuum model. The dip spectrum is plotted
		with a purple diamond, while the flare spectrum uses a green star; all other
		spectra are plotted in black crosses. When plotted versus time, the XIS0
		counting rate is overplotted with a blue dotted line. We overplot the
		best-fit line for the power-law index, cyclotron line energy, and cyclotron
		line width vs.\ luminosity, all of which show significant correlations. The
		reader should note, however, that the power-law index includes points using
		differing continuum models, with the dip spectrum as well as the final two
		spectra adopting a partial-covering absorber and the remaining points using
		a single-absorber model.}
  \label{fig:time}
\end{figure*}

The fitted parameters for the three continuum models are listed in
Table~\ref{tab:time}. We plot selected parameters against time and unabsorbed
flux in Figure~\ref{fig:time}; as the different continuum models show similar
behavior, and in the interests of clarity, we only plot the results from
\texttt{fdcut}. For each plotted parameter versus flux, we perform a linear fit
and calculate 90\% error bars for the slope, as well as computing Pearson's
product-moment correlation coefficient for the data; these results are
summarized in Table~\ref{tab:timecor}.

\begin{deluxetable*}{llrrrrrrrrr}
	\tabletypesize{\footnotesize}
	\tablecolumns{11}
	\tablewidth{0pt}
	\setlength{\tabcolsep}{3pt}
	\tablecaption{Spectral fits for time-resolved spectroscopy}
	\tablehead{ \colhead{} & 
							\colhead{} &
							\colhead{1} &
							\colhead{2} &
							\colhead{3} &
							\colhead{4} &
							\colhead{5} &
							\colhead{6} &
							\colhead{7} &
							\colhead{8} &
							\colhead{9} }
	\startdata
\sidehead{\texttt{highecut}}
	Flux \tablenotemark{a}                  & $\times 10^{-10}$\,erg\,cm$^{-2}$\,s$^{-1}$ & $4.29 \pm  0.03$         & $5.48 \pm  0.05$         & $4.59 \pm  0.05$         & $4.20 \pm  0.05$         & $4.4 \pm  0.2$           & $8.7 \pm  0.1$           & $4.65 \pm  0.05$         & $2.79^{+ 0.10}_{- 0.09}$ & $3.6 \pm  0.1$  \\
	\nh                                     & $\times 10^{22}$\,cm$^{-2}$                 & $2.22 \pm  0.03$         & $2.14 \pm  0.04$         & $2.27 \pm  0.05$         & $1.98 \pm  0.05$         & $3.3 \pm  0.2$           & $2.82 \pm  0.08$         & $2.16 \pm  0.05$         & $2.6 \pm  0.2$           & $3.1 \pm  0.2$  \\
	$N_{\rm H,pcf}$                         & $\times 10^{22}$\,cm$^{-2}$                 & \nodata                  & \nodata                  & \nodata                  & \nodata                  & $17 \pm 3$               & \nodata                  & \nodata                  & $8^{+4}_{-2}$            & $9^{+4}_{-3}$  \\
	$f_{\rm pcf}$                               &                                             & \nodata                  & \nodata                  & \nodata                  & \nodata                  & $0.61 \pm  0.03$         & \nodata                  & \nodata                  & $0.40 \pm  0.09$         & $0.41^{+ 0.09}_{- 0.08}$  \\
	$\Gamma$                                &                                             & $1.16 \pm  0.01$         & $1.18 \pm  0.02$         & $1.21 \pm  0.02$         & $1.20 \pm  0.02$         & $1.13 \pm  0.06$         & $0.94 \pm  0.03$         & $1.11 \pm  0.02$         & $1.38^{+ 0.06}_{- 0.07}$ & $1.38 \pm  0.06$  \\
	$A$ \tablenotemark{b}                   & $\times 10^{-2}$\,ph\,cm$^{-2}$\,s$^{-1}$   & $4.4 \pm  0.1$           & $5.8 \pm  0.2$           & $5.1 \pm  0.2$           & $4.6 \pm  0.2$           & $4.2^{+ 0.6}_{- 0.5}$    & $4.4 \pm  0.1$           & $4.3 \pm  0.2$           & $4.1^{+ 0.6}_{- 0.5}$    & $5.3^{+ 0.7}_{- 0.6}$  \\
	$E_{\mathrm{cut}}$                      & keV                                         & $24^{+2}_{-4}$           & $23^{+3}_{-4}$           & $15^{+3}_{-2}$           & $22 \pm 5$               & $21^{+4}_{-5}$           & $14.1 \pm  0.8$          & $18^{+8}_{-5}$           & $17^{+6}_{-2}$           & $17^{+8}_{-2}$  \\
	$E_{\mathrm{fold}}$                     & keV                                         & $6^{+2}_{-1}$            & $7^{+3}_{-2}$            & $13^{+2}_{-3}$           & $9^{+4}_{-3}$            & $8^{+4}_{-2}$            & $12 \pm 1$               & $11^{+4}_{-5}$           & $12^{+3}_{-5}$           & $12^{+2}_{-6}$  \\
	$E_{\mathrm{K}\alpha}$                  & keV                                         & $6.43 \pm  0.01$         & $6.43 \pm  0.02$         & $6.44 \pm  0.02$         & $6.40 \pm  0.02$         & $6.41 \pm  0.01$         & $6.39 \pm  0.02$         & $6.40 \pm  0.02$         & $6.42 \pm  0.04$         & $6.43 \pm  0.03$  \\
	$I_{\mathrm{K}\alpha}$\tablenotemark{c} & $\times 10^{-4}$\,ph cm$^{-2}$\,s$^{-1}$    & $3.6^{+ 0.4}_{- 0.3}$    & $4.4 \pm  0.7$           & $3.8 \pm  0.6$           & $3.3 \pm  0.6$           & $4.9 \pm  0.5$           & $8 \pm 1$                & $3.2 \pm  0.6$           & $1.2 \pm  0.5$           & $1.9 \pm  0.5$  \\
	$I_{\mathrm{K}\beta}$\tablenotemark{d}  & $\times 10^{-5}$\,ph cm$^{-2}$\,s$^{-1}$    & $4 \pm 3$                & $<8.99$                  & $<9.39$                  & $<7.86$                  & $9 \pm 5$                & $<15.79$                 & $<12.43$                 & $< 4.12$                 & $8 \pm 5$  \\
	$E_{\mathrm{cyc}}$                      & keV                                         & $23 \pm 1$               & $23 \pm 1$               & $21^{+3}_{-1}$           & $22^{+2}_{-1}$           & $21.7^{+1.0}_{- 0.9}$    & $23 \pm 1$               & $21.3^{+1.0}_{- 0.7}$    & $21.4^{+ 0.7}_{- 0.6}$   & $21.8^{+1.0}_{- 0.9}$  \\
	$\sigma_{\mathrm{cyc}}$                 & keV                                         & $3.5 \pm  0.6$           & $2.5^{+ 0.7}_{- 0.5}$    & $3^{+4}_{-2}$            & $2.8^{+1.0}_{- 0.5}$     & $2.5^{+ 0.7}_{- 0.6}$    & $>2.12$                  & $3.2^{+ 0.8}_{- 0.6}$    & $0.8^{+1.0}_{- 0.6}$     & $<3.14$  \\
	$\tau_{\mathrm{cyc}}$                   &                                             & $1.1^{+ 0.1}_{- 0.3}$    & $0.9^{+ 0.2}_{- 0.3}$    & $0.3^{+ 0.2}_{- 0.1}$    & $0.9^{+ 0.2}_{- 0.4}$    & $0.8^{+ 0.2}_{- 0.4}$    & $0.3 \pm  0.1$           & $0.6^{+ 0.4}_{- 0.3}$    & $>0.59$                  & $0.5^{+90.0}_{- 0.2}$  \\
	$\chi^{2}_{\rm red}$(dof)               &                                             & 1.02 (434)               & 1.04 (422)               & 1.09 (423)               & 1.10 (422)               & 1.07 (425)               & 1.08 (421)               & 1.05 (425)               & 1.19 (413)               & 1.06 (419) \\[1ex]
	\hline
	\sidehead{\texttt{fdcut}}
	Flux                                    & $\times 10^{-10}$\,erg\,cm$^{-2}$\,s$^{-1}$ & $4.28 \pm  0.03$         & $5.48 \pm  0.05$         & $4.57^{+ 0.03}_{- 0.05}$ & $4.19 \pm  0.05$         & $4.3 \pm  0.2$           & $8.7 \pm  0.1$           & $4.62 \pm  0.05$         & $2.76^{+ 0.10}_{- 0.09}$ & $3.6 \pm  0.1$  \\
	\nh                                     & $\times 10^{22}$\,cm$^{-2}$                 & $2.21 \pm  0.03$         & $2.14 \pm  0.04$         & $2.22^{+ 0.06}_{- 0.07}$ & $1.95^{+ 0.05}_{- 0.06}$ & $3.2 \pm  0.2$           & $2.7 \pm  0.1$           & $2.05^{+ 0.07}_{- 0.08}$ & $2.5 \pm  0.2$           & $3.1 \pm  0.2$  \\
	$N_{\rm H,pcf}$                         & $\times 10^{22}$\,cm$^{-2}$                 & \nodata                  & \nodata                  & \nodata                  & \nodata                  & $16 \pm 3$               & \nodata                  & \nodata                  & $7^{+4}_{-3}$            & $8^{+4}_{-2}$  \\
	$f_{\rm pcf}$                               &                                             & \nodata                  & \nodata                  & \nodata                  & \nodata                  & $0.59 \pm  0.03$         & \nodata                  & \nodata                  & $0.4 \pm  0.1$           & $0.40 \pm  0.09$  \\
	$\Gamma$                                &                                             & $1.15 \pm  0.02$         & $1.17 \pm  0.02$         & $1.15^{+ 0.05}_{- 0.06}$ & $1.17 \pm  0.04$         & $1.07^{+ 0.07}_{- 0.09}$ & $0.85^{+ 0.06}_{- 0.07}$ & $0.99^{+ 0.06}_{- 0.07}$ & $1.32^{+ 0.08}_{- 0.10}$ & $1.34^{+ 0.07}_{- 0.08}$  \\
	$A$                                     & $\times 10^{-2}$\,ph\,cm$^{-2}$\,s$^{-1}$   & $4.4 \pm  0.1$           & $5.7 \pm  0.2$           & $5.0 \pm  0.2$           & $4.5 \pm  0.2$           & $3.9^{+ 0.6}_{- 0.5}$    & $4.4 \pm  0.1$           & $4.4^{+ 0.4}_{- 0.2}$    & $3.7^{+ 0.6}_{- 0.5}$    & $5.0^{+ 0.7}_{- 0.6}$  \\
	$E_{\mathrm{cut}}$                      & keV                                         & $27 \pm 2$               & $28 \pm 1$               & $24^{+6}_{-5}$           & $27 \pm 3$               & $26^{+2}_{-3}$           & $22^{+5}_{-4}$           & $19^{+5}_{-6}$           & $26^{+2}_{-3}$           & $28 \pm 3$  \\
	$E_{\mathrm{fold}}$                     & keV                                         & $5 \pm 1$                & $4.0^{+1.0}_{- 0.9}$     & $7 \pm 2$                & $6 \pm 2$                & $5 \pm 1$                & $7^{+1}_{-2}$            & $9 \pm 2$                & $5 \pm 2$                & $5 \pm 2$  \\
	$E_{\mathrm{K}\alpha}$                  & keV                                         & $6.43 \pm  0.01$         & $6.43 \pm  0.02$         & $6.44 \pm  0.02$         & $6.40 \pm  0.02$         & $6.41 \pm  0.01$         & $6.39 \pm  0.02$         & $6.40 \pm  0.02$         & $6.42 \pm  0.04$         & $6.42 \pm  0.03$  \\
	$I_{\mathrm{K}\alpha}$                  & $\times 10^{-4}$\,ph cm$^{-2}$\,s$^{-1}$    & $3.5 \pm  0.3$           & $4.4 \pm  0.7$           & $3.7 \pm  0.6$           & $3.3 \pm  0.6$           & $4.9 \pm  0.5$           & $7 \pm 1$                & $3.0 \pm  0.6$           & $1.2^{+ 0.5}_{- 0.4}$    & $1.9 \pm  0.5$  \\
	$I_{\mathrm{K}\beta}$                   & $\times 10^{-5}$\,ph cm$^{-2}$\,s$^{-1}$    & $4 \pm 3$                & $<8.76$                  & $<9.03$                  & $<7.58$                  & $9 \pm 5$                & $<14.20$                 & $<11.33$                 & $< 3.69$                 & $8 \pm 5$  \\
	$E_{\mathrm{cyc}}$                      & keV                                         & $23.2^{+ 0.7}_{- 0.5}$   & $23.2^{+ 0.8}_{- 0.6}$   & $23^{+2}_{-1}$           & $22.6^{+ 0.8}_{- 0.6}$   & $22.3^{+ 0.6}_{- 0.5}$   & $24^{+2}_{-1}$           & $22.1^{+ 0.8}_{- 0.7}$   & $22.0^{+ 0.8}_{- 0.6}$   & $23 \pm 1$  \\
	$\sigma_{\mathrm{cyc}}$                 & keV                                         & $3.4 \pm  0.6$           & $2.7 \pm  0.7$           & $3^{+2}_{-1}$            & $2.4^{+1.0}_{- 0.8}$     & $2.4^{+ 0.8}_{- 0.7}$    & $3^{+2}_{-1}$            & $2 \pm 1$                & $<2.41$                  & $3 \pm 1$  \\
	$\tau_{\mathrm{cyc}}$                   &                                             & $0.9^{+ 0.2}_{- 0.1}$    & $0.8 \pm  0.1$           & $0.4^{+ 0.2}_{- 0.1}$    & $0.6^{+ 0.2}_{- 0.1}$    & $0.7 \pm  0.1$           & $0.5^{+ 0.2}_{- 0.1}$    & $0.5^{+ 0.6}_{- 0.1}$    & $>0.74$                  & $0.7 \pm  0.2$  \\
	$\chi^{2}_{\rm red}$(dof)               &                                             & 1.04 (434)               & 1.06 (422)               & 1.11 (423)               & 1.10 (422)               & 1.09 (425)               & 1.08 (420)               & 1.03 (425)               & 1.20 (413)               & 1.07 (419) \\[1ex]
	\hline
	\sidehead{\texttt{npex}}
	Flux                                    & $\times 10^{-10}$\,erg\,cm$^{-2}$\,s$^{-1}$ & $4.24 \pm  0.03$         & $5.51 \pm  0.06$         & $4.51 \pm  0.05$         & $4.15 \pm  0.05$         & $4.7^{+ 0.4}_{- 0.3}$    & $8.6 \pm  0.1$           & $4.55 \pm  0.05$         & $2.8 \pm  0.2$           & $4.0^{+ 0.5}_{- 0.3}$  \\
	\nh                                     & $\times 10^{22}$\,cm$^{-2}$                 & $2.05 \pm  0.05$         & $2.14 \pm  0.07$         & $2.03^{+ 0.08}_{- 0.07}$ & $1.76 \pm  0.07$         & $3.4 \pm  0.3$           & $2.5 \pm  0.1$           & $1.83 \pm  0.07$         & $2.5 \pm  0.2$           & $3.2 \pm  0.3$  \\
	$N_{\rm H,pcf}$                         & $\times 10^{22}$\,cm$^{-2}$                 & \nodata                  & \nodata                  & \nodata                  & \nodata                  & $19 \pm 3$               & \nodata                  & \nodata                  & $10 \pm 4$               & $12^{+4}_{-3}$  \\
	$f_{\rm pcf}$                               &                                             & \nodata                  & \nodata                  & \nodata                  & \nodata                  & $0.66^{+ 0.05}_{- 0.06}$ & \nodata                  & \nodata                  & $0.4^{+ 0.1}_{- 0.2}$    & $0.5 \pm  0.1$  \\
	$\alpha$                                &                                             & $0.63^{+ 0.04}_{- 0.03}$ & $0.79^{+ 0.06}_{- 0.05}$ & $0.63^{+ 0.06}_{- 0.05}$ & $0.63 \pm  0.05$         & $0.8 \pm  0.2$           & $0.36^{+ 0.07}_{- 0.06}$ & $0.48^{+ 0.05}_{- 0.04}$ & $0.8 \pm  0.2$           & $1.1 \pm  0.3$  \\
	$A$                                     & $\times 10^{-2}$\,ph\,cm$^{-2}$\,s$^{-1}$   & $4.3 \pm  0.2$           & $6.7 \pm  0.5$           & $4.5^{+ 0.4}_{- 0.3}$    & $4.1 \pm  0.3$           & $6^{+4}_{-2}$            & $4.3 \pm  0.2$           & $3.5^{+ 0.3}_{- 0.2}$    & $4^{+2}_{-1}$            & $8^{+7}_{-3}$  \\
	$B$                                     & $\times 10^{-3}$                            & $3.2 \pm  0.3$           & $3.4 \pm  0.3$           & $2.4 \pm  0.4$           & $2.5 \pm  0.4$           & $2.8 \pm  0.7$           & $4^{+1}_{-2}$            & $2.4 \pm  0.4$           & $2.0^{+ 0.5}_{- 0.4}$    & $1.5 \pm  0.5$  \\
	$k_{\rm B}T$                                    & keV                                         & $4.8 \pm  0.1$           & $4.7 \pm  0.1$           & $5.2^{+ 0.4}_{- 0.2}$    & $5.1 \pm  0.2$           & $4.8 \pm  0.2$           & $5.4^{+1.0}_{- 0.4}$     & $5.5^{+ 0.3}_{- 0.2}$    & $4.7 \pm  0.3$           & $4.7 \pm  0.3$  \\
	$E_{\mathrm{K}\alpha}$                  & keV                                         & $6.43 \pm  0.01$         & $6.43 \pm  0.02$         & $6.44 \pm  0.02$         & $6.40 \pm  0.02$         & $6.41 \pm  0.01$         & $6.39 \pm  0.02$         & $6.40 \pm  0.02$         & $6.42 \pm  0.04$         & $6.42 \pm  0.03$  \\
	$I_{\mathrm{K}\alpha}$                  & $\times 10^{-4}$\,ph cm$^{-2}$\,s$^{-1}$    & $3.6^{+ 0.4}_{- 0.3}$    & $4.6 \pm  0.7$           & $3.8 \pm  0.6$           & $3.4 \pm  0.6$           & $5.1 \pm  0.6$           & $8 \pm 1$                & $3.1 \pm  0.6$           & $1.2 \pm  0.5$           & $2.1 \pm  0.6$  \\
	$I_{\mathrm{K}\beta}$                   & $\times 10^{-5}$\,ph cm$^{-2}$\,s$^{-1}$    & $4 \pm 3$                & $<8.80$                  & $<10.05$                 & $<8.45$                  & $10 \pm 5$               & $<14.67$                 & $<12.49$                 & $< 4.18$                 & $10^{+6}_{-5}$  \\
	$E_{\mathrm{cyc}}$                      & keV                                         & $22.6 \pm  0.5$          & $22.3 \pm  0.6$          & $23^{+2}_{-1}$           & $22.3^{+ 0.7}_{- 0.6}$   & $21.9 \pm  0.5$          & $25^{+4}_{-2}$           & $22.3^{+ 0.8}_{- 0.7}$   & $21.8^{+ 0.7}_{- 0.6}$   & $22.6^{+1.0}_{- 0.9}$  \\
	$\sigma_{\mathrm{cyc}}$                 & keV                                         & $2.7 \pm  0.5$           & $2.4 \pm  0.7$           & $3^{+2}_{-1}$            & $2.3^{+ 0.8}_{- 0.7}$    & $2.3 \pm  0.7$           & $3^{+3}_{-2}$            & $2.0 \pm  0.9$           & $1.0^{+1.0}_{- 0.7}$     & $3 \pm 1$  \\
	$\tau_{\mathrm{cyc}}$                   &                                             & $0.66^{+ 0.09}_{- 0.08}$ & $0.6 \pm  0.1$           & $0.4 \pm  0.1$           & $0.6^{+ 0.2}_{- 0.1}$    & $0.6 \pm  0.1$           & $0.5^{+ 0.4}_{- 0.1}$    & $0.5^{+ 0.3}_{- 0.1}$    & $>0.69$                  & $0.6^{+ 0.2}_{- 0.1}$  \\
	$\chi^{2}_{\rm red}$(dof)               &                                             & 1.08 (434)               & 1.00 (422)               & 1.12 (423)               & 1.11 (422)               & 1.08 (425)               & 1.07 (420)               & 1.01 (425)               & 1.20 (413)               & 1.06 (419)
	\enddata
	\tablenotetext{a}{2--10\,keV unabsorbed flux}
	\tablenotetext{b}{Power-law normalization at 1 keV}
	\tablenotetext{c}{Iron K$\alpha$ intensity}
	\tablenotetext{d}{Iron K$\beta$ intensity}
	\label{tab:time}
\end{deluxetable*}

\begin{deluxetable*}{llrrrrrrrrr}
  \tabletypesize{\scriptsize}
  \tablecolumns{8}
  \tablewidth{0pt}
  \tablecaption{Linear fits (with 90\% error bars) and correlation coefficients for time-resolved spectroscopy}
  \tablehead{ \colhead{} & 
              \colhead{} &
              \multicolumn{3}{c}{\texttt{highecut}} &
              \multicolumn{3}{c}{\texttt{fdcut}} &
              \multicolumn{3}{c}{\texttt{npex}} \\
              \colhead{Parameter} &
              \colhead{Units} &
              \colhead{Slope \tablenotemark{a}} &
              \colhead{$r$ \tablenotemark{b}} &
							\colhead{$p$-value \tablenotemark{c}} &
              \colhead{Slope} &
              \colhead{$r$} &
							\colhead{$p$-value} &
              \colhead{Slope} &
              \colhead{$r$} &
							\colhead{$p$-value}}
  \startdata
  $\Gamma$                &     & $-0.054^{+ 0.006}_{- 0.006}$ & $-0.868$ & $0.002$		& $-0.05^{+ 0.02}_{- 0.02}$ & $-0.846$ &	$0.004$	& \nodata                   & \nodata		&	\nodata \\
  $E_{\mathrm{cut}}$      & keV & $-1.3^{+ 0.4}_{- 0.4}$       & $-0.369$ & $> 0.1$		& $-0.5^{+ 0.9}_{- 0.9}$    & $-0.465$ &	$> 0.1$	& \nodata                   & \nodata 	&	\nodata \\
  $E_{\mathrm{fold}}$     & keV & $0.6^{+ 0.3}_{- 0.3}$        & $0.078$  & $> 0.1$		& $0.2^{+ 0.3}_{- 0.3}$     & $0.327$  &	$> 0.1$	& \nodata                   & \nodata 	&	\nodata \\
  $\alpha$                &     & \nodata                      & \nodata  & \nodata		& \nodata                   & \nodata  &	\nodata	& $-0.05^{+ 0.02}_{- 0.02}$ & $-0.689$	&	$0.040$	\\
  $k_{\rm B}T$            & keV & \nodata                      & \nodata  & \nodata		& \nodata                   & \nodata  &	\nodata	& $0.06^{+ 0.06}_{- 0.06}$  & $0.599$ 	&	$0.088$	\\
  $E_{\mathrm{cyc}}$      & keV & $0.3^{+ 0.2}_{- 0.2}$        & $0.586$  & $0.097$		& $0.3^{+ 0.2}_{- 0.2}$     & $0.690$  &	$0.040$	& $0.4^{+ 0.2}_{- 0.2}$     & $0.896$ 	&	$0.001$	\\
  $\sigma_{\mathrm{cyc}}$ & keV & $0.1^{+ 0.5}_{- 0.6}$        & $0.094$  & $> 0.1$		& $0.2^{+ 0.3}_{- 0.3}$     & $0.407$  &	$> 0.1$	& $0.3^{+ 0.3}_{- 0.3}$     & $0.624$ 	&	$0.072$	\\
  $\tau_{\mathrm{cyc}}$   &     & $-0.12^{+ 0.04}_{- 0.04}$    & $-0.340$ & $> 0.1$		& $-0.03^{+ 0.04}_{- 0.04}$ & $-0.422$ &	$> 0.1$	& $-0.02^{+ 0.03}_{- 0.03}$ & $-0.534$	&	$> 0.1$
  \enddata
  \tablenotetext{a}{From linear fit against unabsorbed 2--10\,keV flux in units of $10^{-10}$\,erg\,cm$^{-2}$\,s$^{-1}$}
  \tablenotetext{b}{Pearson product-moment correlation coefficient}
	\tablenotetext{c}{Two-tailed $p$-value for $N=9$}
  \label{tab:timecor}
\end{deluxetable*}

\subsection{Pulse-to-pulse analysis} \label{ssec:pulse}

\begin{figure*}[ht]
	\centering
	\epsscale{1.0}
	\plottwo{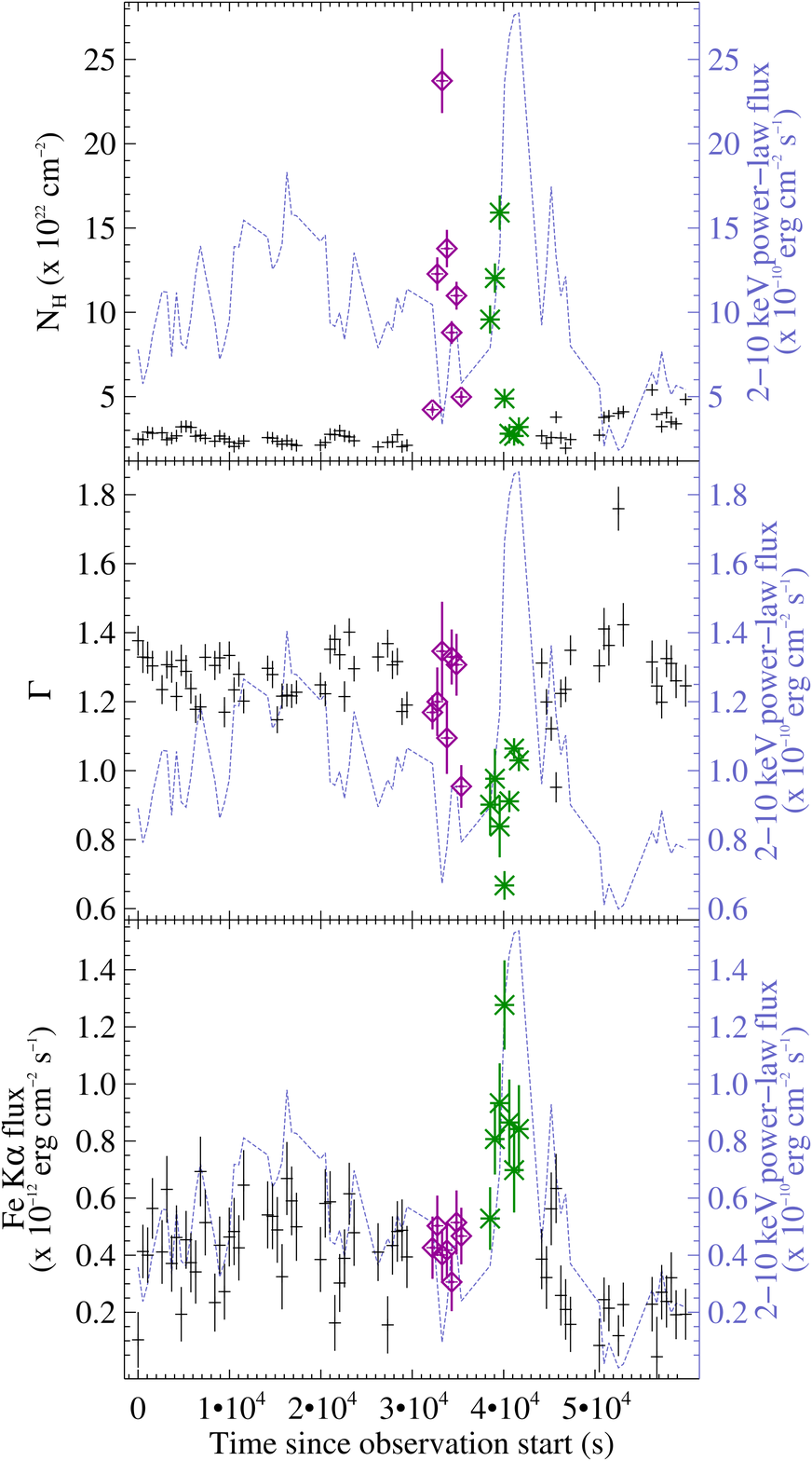}{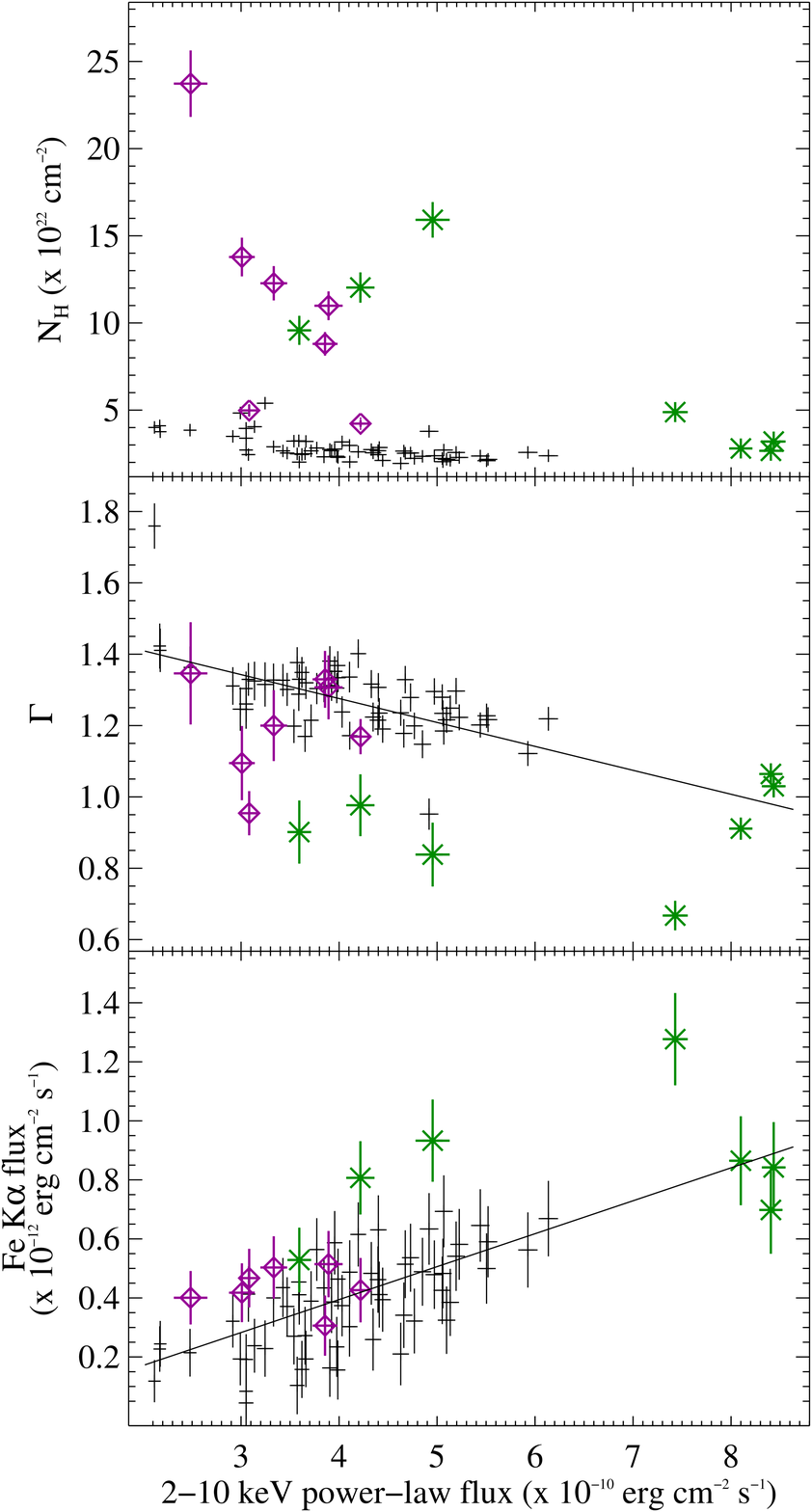}
	\caption{Parameters versus time (left) and versus flux (right) for the
		pulse-to-pulse analysis. We plot the first spike in \nh in purple diamonds
		and the approximate extent of the ``flare'' state of the source in green
		stars, with the remainder of the data in black crosses. The unabsorbed
		power-law flux is rescaled and overplotted as a dashed blue line on the
		plots versus time. Correlations with luminosity are apparent for the
		power-law index and the iron K$\alpha$ flux, while the absorbing column
		density reveals that the dip at around 32\,ks in the XIS lightcurve is due
		to two large spikes in \nh.}
  \label{fig:pulse}
\end{figure*}

To obtain a more complete picture of the variability of the absorbing column and
the iron line, we further divided the observation and extracted spectra for
individual pulses. We only examine spectra for full pulses - spectra from
fractional pulses are excluded, to avoid contamination by the significant
spectral variability within each pulse (see Section~\ref{ssec:phase}). After
excluding fractional pulses, we have a total of 76 spectra. Only XIS spectra are
analyzed, as the low exposure for each spectrum precludes the use of PIN data.
Each spectrum is fit with a simple model consisting of a single absorbed
power-law with an additive Gaussian modeling the iron K$\alpha$ line. This model
fits the spectra generally quite well, with an average reduced $\chi^{2}$ of
$1.17$. No evidence for partial covering was visible in the residuals, and the
partial-covering model used in the previous section for the dip could not be
constrained, so the absorption was modeled by a single \texttt{tbnew} component.
The \texttt{cflux} model component in XSPEC was used to obtain unabsorbed fluxes
for the power-law and the iron line. We plot each parameter with respect to time
and luminosity in Figure~\ref{fig:pulse}. We also calculated Pearson's
correlation coefficients and performed linear fits on selected parameter versus
the 2--10\,keV power-law flux; these results are summarized in
Table~\ref{tab:pulsecor}. The measured power-law index behaves roughly similar
to what was seen in the time-resolved spectroscopy in Section \ref{ssec:time},
despite the use of a partial-covering model in that analysis. However, while the
measured absorbing column density in the dip lines up roughly with the
partial-covering \nh found in the time-resolved spectroscopy, the
slightly-elevated \nh in the final $\sim 10$ pulse-to-pulse spectra do not reach
the same levels as the partial-covering model found for the final two
time-resolved spectra.

\begin{deluxetable}{llrrr}
  \tabletypesize{\scriptsize}
  \tablecolumns{4}
  \tablewidth{0pt}
  \tablecaption{Linear fits and correlation coefficients relative to 2--10\,keV power-law flux for pulse-to-pulse spectroscopy}
  \tablehead{ \colhead{Parameter} &
              \colhead{Units} &
              \colhead{Slope \tablenotemark{a}} &
              \colhead{$r$ \tablenotemark{b}} &
							\colhead{$p$-value \tablenotemark{c}}}
  \startdata
  \nh                    & $10^{22}$\,cm$^{-2}$                  &  $-0.07 \pm 0.03$ & $-0.228$ & $0.05$ \\
	$\Gamma$        & &  $-0.067 \pm 0.008$ & $-0.570$ & $10^{-7}$ \\
	Fe K$\alpha$ flux & $10^{-10}$\,erg\,cm$^{-2}$\,s$^{-1}$  &  $0.011 \pm 0.002$ & $0.707$ & $< 10^{-8}$ \\
  \enddata
  \tablenotetext{a}{From linear fit against 2--10\,keV unabsorbed flux in units of $10^{-10}$\,erg\,cm$^{-2}$\,s$^{-1}$}
  \tablenotetext{b}{Pearson product-moment correlation coefficient}
	\tablenotetext{c}{Two-tailed $p$-value for $N=75$}
  \label{tab:pulsecor}
\end{deluxetable}

\subsection{Phase-resolved spectroscopy}
\label{ssec:phase}

\begin{figure*}[ht]
	\centering
  \plotone{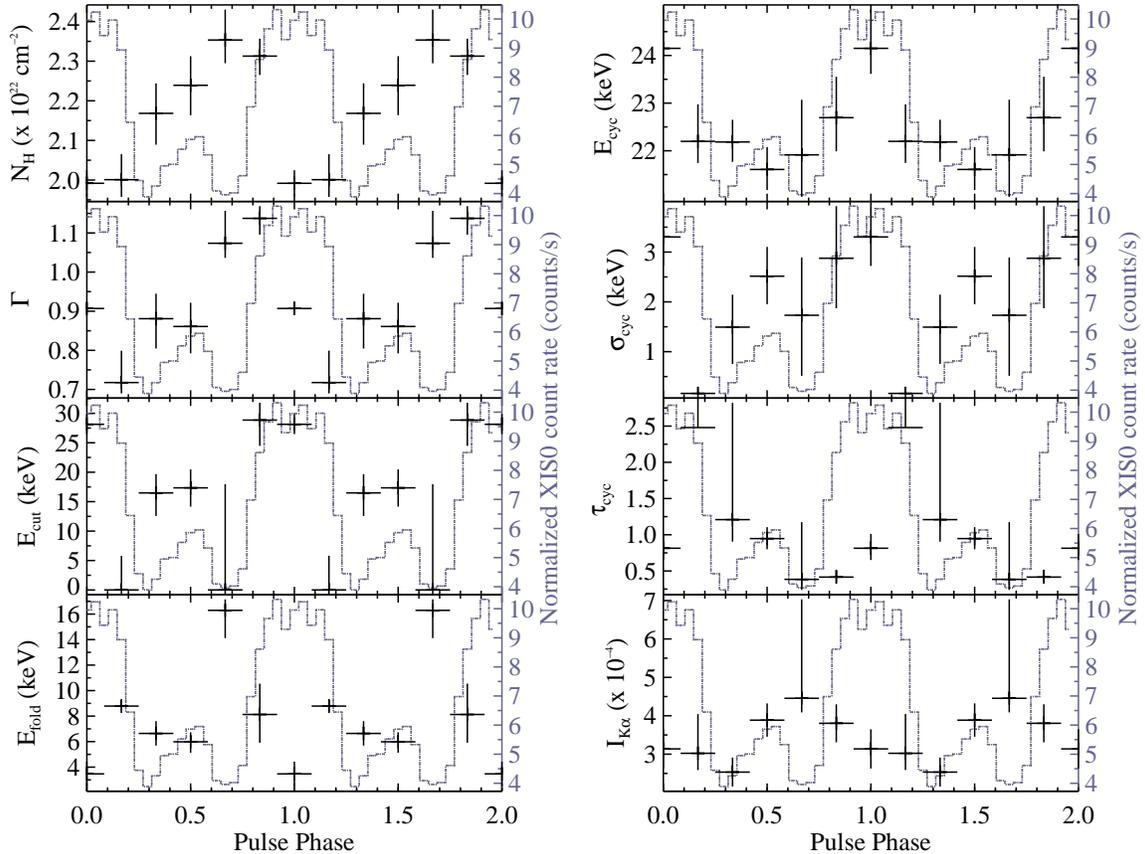}
	\caption{Selected parameters for the phase-resolved fits. The spectral
		parameters for the \texttt{fdcut} continuum are plotted as crosses, while
		the pulse profile in the 2-10\,keV band is overplotted with a blue dotted
		line.  For clarity, the pulse profile is repeated. The vertical axis on the
		right of each plot indicates the normalized XIS count rate for the pulse
		profile.  The apparent dependence of \nh on pulse phase is dependent on the
		choice of continuum model and likely reflects the fact that our empirical
		continuum models do not perfectly model the underlying continuum, rather than
		representing a true modulation of the obscuring column density.}
  \label{fig:phase}
\end{figure*}

The pulse profile of \fif is double-peaked, with a large primary pulse and
a significantly smaller secondary pulse (see Figure~\ref{fig:pulse}), similar to
that found by \citet{robba_bepposax_2001}, \citet{rodes-roca_detecting_2010},
and \citet{hemphill_2013}. Our phase-resolved analysis was carried out by
defining good time intervals for six phase bins, with phase zero defined as the
peak of the main pulse. The remaining phase bins thus cover the rising and
falling edges of the main pulse, the peak of the secondary pulse, and the
low-flux dips on either side of the secondary pulse. This choice of phase
binning covers all the major features of the pulse profile without sacrificing
signal-to-noise. XIS and PIN spectra were analyzed simultaneously, using the
three models described in Section~\ref{ssec:time}. The parameters of these three
models with 90\% error bars are listed in Table~\ref{tab:phase}, and the
parameters for the \texttt{fdcut} continuum are plotted in
Figure~\ref{fig:phase}. The shape of the continuum, including the cyclotron
line, changes considerably between the rising and falling edges of the main
pulse. Interestingly, the absorbing column density appears to be significantly
dependent on pulse phase (although the magnitude of the effect is relatively
small, on the order of 10\%).  A physical mechanism for this effect is difficult
to imagine, and on the whole unlikely - the absorber does not readily show signs
of significant ionization, which one would expect if it was close enough to the
neutron star to be effected by the star's rotation.  It is likely that the
dependence of \nh on pulse phase is in part due to our unphysical modeling of
the continuum.  This explanation is favored by the fact that the dependence of
\nh on pulse phase varies with the choice of continuum model - under
\texttt{highecut}, \nh is roughly constant in all bins save the peak of the main
pulse, where it drops suddenly, while in \texttt{fdcut} and \texttt{npex} the
\nh follows a more ``sawtooth'' pattern, dropping sharply at the main pulse but
rising slowly afterwards. Meanwhile, the CRSF parameters and iron line intensity
all show strong dependence on phase, without any strong dependence on continuum
model.

\begin{deluxetable*}{llrrrrrr}
  \tabletypesize{\scriptsize}
  \tablecolumns{8}
  \tablewidth{0pt}
  \tablecaption{Spectral fits for phase-resolved spectroscopy}
  \tablehead{ \colhead{} & 
              \colhead{} &
              \multicolumn{6}{c}{Phase bin} \\
              \colhead{} &
              \colhead{} &
              \multicolumn{1}{c}{1} &
              \multicolumn{1}{c}{2} &
              \multicolumn{1}{c}{3} &
              \multicolumn{1}{c}{4} &
              \multicolumn{1}{c}{5} &
              \multicolumn{1}{c}{6} }
  \startdata
 \sidehead{\texttt{highecut}}
  Flux\tablenotemark{a}                   & $\times 10^{-10}$\,erg\,cm$^{-2}$\,s$^{-1}$ & $6.32 \pm  0.04$         & $4.59 \pm  0.04$         & $2.98 \pm  0.03$         & $3.67 \pm  0.03$         & $2.60 \pm  0.03$            & $5.47 \pm  0.04$  \\
  \nh                                     & $\times 10^{22}$\,cm$^{-2}$                 & $2.00 \pm  0.03$         & $2.34 \pm  0.04$         & $2.30 \pm  0.05$         & $2.35 \pm  0.05$         & $2.53 \pm  0.05$            & $2.35 \pm  0.04$  \\
  $\Gamma$                                &                                             & $0.91 \pm  0.01$         & $1.13 \pm  0.02$         & $1.02 \pm  0.02$         & $0.98 \pm  0.02$         & $1.28 \pm  0.02$            & $1.18 \pm  0.02$  \\
  $A$\tablenotemark{b}                    & $\times 10^{-2}$\,ph\,cm$^{-2}$\,s$^{-1}$   & $4.2 \pm  0.1$           & $4.4 \pm  0.1$           & $2.37 \pm  0.09$         & $2.72 \pm  0.09$         & $3.2 \pm  0.1$              & $5.7^{+ 0.2}_{- 0.1}$  \\
  $E_{\mathrm{cut}}$                      & keV                                         & $23 \pm 3$               & $12.5^{+ 0.6}_{- 0.7}$   & $14^{+4}_{-1}$           & $23^{+4}_{-8}$           & $14^{+9}_{-2}$              & $18^{+8}_{-4}$  \\
  $E_{\mathrm{fold}}$                     & keV                                         & $6^{+1}_{-2}$            & $10.1 \pm  0.7$          & $9.1^{+ 0.9}_{-2.0}$     & $4^{+4}_{-1}$            & $21^{+5}_{-10}$             & $17^{+5}_{-7}$  \\
  $E_{\mathrm{K}\alpha}$                  & keV                                         & $6.41 \pm  0.02$         & $6.41 \pm  0.02$         & $6.42 \pm  0.02$         & $6.43 \pm  0.01$         & $6.415^{+ 0.009}_{- 0.008}$ & $6.42 \pm  0.01$  \\
  $I_{\mathrm{K}\alpha}$\tablenotemark{c} & $\times 10^{-4}$\,ph cm$^{-2}$\,s$^{-1}$    & $3.2 \pm  0.5$           & $3.3 \pm  0.4$           & $2.6 \pm  0.4$           & $4.0 \pm  0.4$           & $4.6 \pm  0.4$              & $3.9 \pm  0.5$  \\
  $I_{\mathrm{K}\beta}$\tablenotemark{d}  & $\times 10^{-5}$\,ph cm$^{-2}$\,s$^{-1}$    & $<8.00$                  & $<7.00$                  & $5 \pm 4$                & $5 \pm 4$                & $6 \pm 3$                   & $<7.00$  \\
  $E_{\mathrm{cyc}}$                      & keV                                         & $23.4^{+1.0}_{- 0.8}$    & $21.9 \pm  0.6$          & $21.9 \pm  0.5$          & $22 \pm 2$               & $21 \pm 1$                  & $21.4^{+ 0.7}_{- 0.6}$  \\
  $\sigma_{\mathrm{cyc}}$                 & keV                                         & $3.0^{+ 0.6}_{- 0.3}$    & $0.33$ (fixed)           & $2 \pm 1$                & $3.5^{+1.0}_{- 0.4}$     & $2.82$ (fixed)              & $3.0^{+ 0.6}_{-2.0}$  \\
  $\tau_{\mathrm{cyc}}$                   &                                             & $0.9 \pm  0.2$           & $2^{+2}_{-1}$            & $1.0^{+2.0}_{- 0.3}$     & $1.7^{+ 0.2}_{- 0.9}$    & $0.3^{+ 0.4}_{- 0.1}$       & $0.4^{+ 0.3}_{- 0.2}$  \\
  $\chi^{2}_{\rm red}$(dof)               &                                             & 1.14 (432)               & 1.36 (423)               & 1.28 (417)               & 1.08 (423)               & 1.14 (427)                  & 1.17 (438) \\[1ex]
  \hline
  \sidehead{\texttt{fdcut}}
  Flux                                    & $\times 10^{-10}$\,erg\,cm$^{-2}$\,s$^{-1}$ & $6.32 \pm  0.04$         & $4.49^{+ 0.04}_{- 0.03}$ & $2.94 \pm  0.03$         & $3.63 \pm  0.04$         & $2.58 \pm  0.03$            & $5.46 \pm  0.04$  \\
  \nh                                     & $\times 10^{22}$\,cm$^{-2}$                 & $1.99 \pm  0.03$         & $2.00^{+ 0.06}_{- 0.04}$ & $2.17 \pm  0.08$         & $2.24^{+ 0.07}_{- 0.08}$ & $2.35^{+ 0.08}_{- 0.06}$    & $2.31^{+ 0.04}_{- 0.05}$  \\
  $\Gamma$                                &                                             & $0.91^{+ 0.01}_{- 0.02}$ & $0.72^{+ 0.08}_{- 0.03}$ & $0.88^{+ 0.06}_{- 0.08}$ & $0.86^{+ 0.06}_{- 0.07}$ & $1.07^{+ 0.08}_{- 0.04}$    & $1.14^{+ 0.03}_{- 0.04}$  \\
  $A$                                     & $\times 10^{-2}$\,ph\,cm$^{-2}$\,s$^{-1}$   & $4.2 \pm  0.1$           & $6.4^{+ 0.2}_{-1.0}$     & $2.26^{+ 0.10}_{- 0.09}$ & $2.56^{+ 0.10}_{- 0.09}$ & $5.4^{+ 0.3}_{-2.0}$        & $5.7 \pm  0.2$  \\
  $E_{\mathrm{cut}}$                      & keV                                         & $28 \pm 2$               & $<6.01$                  & $16^{+3}_{-4}$           & $17 \pm 3$               & $<20.01$                    & $29^{+3}_{-4}$  \\
  $E_{\mathrm{fold}}$                     & keV                                         & $3 \pm 1$                & $8.8^{+ 0.2}_{- 0.5}$    & $6.6^{+1.0}_{- 0.9}$     & $6.0^{+ 0.7}_{- 0.9}$    & $16.3^{+ 0.9}_{-2.0}$       & $8 \pm 2$  \\
  $E_{\mathrm{K}\alpha}$                  & keV                                         & $6.41 \pm  0.02$         & $6.42 \pm  0.02$         & $6.42 \pm  0.02$         & $6.43 \pm  0.01$         & $6.415 \pm  0.009$          & $6.42^{+ 0.02}_{- 0.01}$  \\
  $I_{\mathrm{K}\alpha}$                  & $\times 10^{-4}$\,ph cm$^{-2}$\,s$^{-1}$    & $3.1 \pm  0.5$           & $3.0^{+1.0}_{- 0.4}$     & $2.5 \pm  0.4$           & $3.9 \pm  0.4$           & $4.5^{+3.0}_{- 0.4}$        & $3.8 \pm  0.5$  \\
  $I_{\mathrm{K}\beta}$                   & $\times 10^{-5}$\,ph cm$^{-2}$\,s$^{-1}$    & $<8.00$                  & $<6.00$                  & $5 \pm 4$                & $<8.00$                  & $6^{+5}_{-3}$               & $<7.00$  \\
  $E_{\mathrm{cyc}}$                      & keV                                         & $24.1^{+ 0.8}_{- 0.5}$   & $22.2^{+ 0.8}_{- 0.5}$   & $22.2^{+ 0.5}_{- 0.4}$   & $21.6^{+ 0.5}_{- 0.4}$   & $21.9^{+1.0}_{- 0.9}$       & $22.7^{+ 0.9}_{- 0.7}$  \\
  $\sigma_{\mathrm{cyc}}$                 & keV                                         & $3.3 \pm  0.6$           & $0.16$ (fixed)           & $1.5 \pm  0.7$           & $2.5 \pm  0.6$           & $2 \pm 1$                   & $3 \pm 1$  \\
  $\tau_{\mathrm{cyc}}$                   &                                             & $0.8 \pm  0.2$           & $> 7.69$                 & $1.2^{+2.0}_{- 0.3}$     & $0.9^{+ 0.2}_{- 0.1}$    & $0.4^{+ 0.8}_{- 0.1}$       & $0.42^{+ 0.10}_{- 0.08}$  \\
  $\chi^{2}_{\rm red}$(dof)               &                                             & 1.15 (432)               & 1.05 (423)               & 1.26 (417)               & 1.05 (423)               & 1.10 (426)                  & 1.18 (438) \\[1ex]
  \hline
  \sidehead{\texttt{npex}}
  Flux                                    & $\times 10^{-10}$\,erg\,cm$^{-2}$\,s$^{-1}$ & $6.31^{+ 0.05}_{- 0.04}$ & $4.42 \pm  0.04$         & $2.90 \pm  0.03$         & $3.60 \pm  0.04$         & $2.54 \pm  0.03$            & $5.40 \pm  0.04$  \\
  \nh                                     & $\times 10^{22}$\,cm$^{-2}$                 & $1.91 \pm  0.05$         & $1.83 \pm  0.05$         & $1.95 \pm  0.08$         & $2.05 \pm  0.08$         & $2.15^{+ 0.07}_{- 0.06}$    & $2.13 \pm  0.05$  \\
  $\alpha$                                &                                             & $0.50 \pm  0.04$         & $0.32 \pm  0.03$         & $0.33 \pm  0.06$         & $0.33 \pm  0.06$         & $0.64 \pm  0.04$            & $0.66 \pm  0.03$  \\
  $A$                                     & $\times 10^{-2}$\,ph\,cm$^{-2}$\,s$^{-1}$   & $4.5^{+ 0.3}_{- 0.2}$    & $3.1^{+ 0.2}_{- 0.1}$    & $1.9 \pm  0.2$           & $2.4 \pm  0.2$           & $2.4^{+ 0.2}_{- 0.1}$       & $5.1 \pm  0.3$  \\
  $B$                                     & $\times 10^{-3}$                            & $5.2 \pm  0.3$           & $2.8 \pm  0.6$           & $4.9 \pm  0.8$           & $6.7 \pm  0.9$           & $1.1^{+ 0.4}_{- 0.3}$       & $1.8 \pm  0.2$  \\
	$k_{\rm B}T$                                    & keV                                         & $5.0 \pm  0.1$           & $4.8 \pm  0.2$           & $4.6 \pm  0.2$           & $4.3 \pm  0.2$           & $6.1^{+ 0.6}_{- 0.4}$       & $5.9 \pm  0.2$  \\
  $E_{\mathrm{K}\alpha}$                  & keV                                         & $6.41 \pm  0.02$         & $6.42 \pm  0.02$         & $6.42 \pm  0.02$         & $6.43 \pm  0.01$         & $6.415^{+ 0.009}_{- 0.008}$ & $6.42 \pm  0.01$  \\
  $I_{\mathrm{K}\alpha}$                  & $\times 10^{-4}$\,ph cm$^{-2}$\,s$^{-1}$    & $3.2 \pm  0.5$           & $3.2 \pm  0.5$           & $2.6 \pm  0.4$           & $3.9 \pm  0.5$           & $4.6 \pm  0.4$              & $4.0 \pm  0.5$  \\
  $I_{\mathrm{K}\beta}$                   & $\times 10^{-5}$\,ph cm$^{-2}$\,s$^{-1}$    & $<7.00$                  & $<8.00$                  & $5 \pm 4$                & $5 \pm 4$                & $7 \pm 3$                   & $<8.00$  \\
  $E_{\mathrm{cyc}}$                      & keV                                         & $23.9^{+ 0.5}_{- 0.4}$   & $22.1^{+ 0.5}_{- 0.4}$   & $22.3^{+ 0.5}_{- 0.4}$   & $21.7^{+ 0.5}_{- 0.4}$   & $21.9^{+1.0}_{- 0.8}$       & $22.4 \pm  0.6$  \\
  $\sigma_{\mathrm{cyc}}$                 & keV                                         & $2.1 \pm  0.5$           & $0.33$ (fixed)           & $1.5^{+ 0.6}_{- 0.7}$    & $2.4 \pm  0.5$           & $2 \pm 1$                   & $3.4 \pm  0.7$  \\
  $\tau_{\mathrm{cyc}}$                   &                                             & $0.48^{+ 0.09}_{- 0.08}$ & $4^{+4}_{-2}$            & $1.2^{+2.0}_{- 0.3}$     & $0.9^{+ 0.2}_{- 0.1}$    & $0.4^{+ 0.2}_{- 0.1}$       & $0.46 \pm  0.06$  \\
  $\chi^{2}_{\rm red}$(dof)               &                                             & 1.24 (432)               & 1.05 (424)               & 1.27 (417)               & 1.07 (423)               & 1.07 (426)                  & 1.15 (438)
  \enddata
  \tablenotetext{a}{2--10\,keV unabsorbed flux}
  \tablenotetext{b}{Power-law normalization at 1 keV}
	\tablenotetext{c}{Iron K$\alpha$ intensity}
	\tablenotetext{d}{Iron K$\beta$ intensity}
  \label{tab:phase}
\end{deluxetable*}

\section{Discussion}
\label{sec:disc}

Our observation of \fif reveals intriguing results in three broad areas. The
line-of-sight column density increases drastically $\sim 32$\,ks into the
observation, and this is followed quickly by an X-ray flare. The increase in \nh
is not associated with any additional changes in the continuum, implying that
the source of the \nh increase is relatively distant from the neutron star
accretion column, while the flare is associated with a significant increase in
spectral hardness. The iron K$\alpha$ intensity varies with luminosity and with
pulse phase, giving an indication of the illuminated iron's overall distance
from the neutron star. And the cyclotron line energy has a weak dependence on
the source's luminosity, revealed by the high-luminosity flare.

\subsection{Variability of Absorption and Continuum}
\label{ssec:abs}

As can be seen in Figures~\ref{fig:time} and \ref{fig:pulse}, the column density
and continuum parameters during the pre-dip portion of the observation stay
relatively constant, with \nh at an average value of $2.16 \times
10^{22}$\,cm$^{-2}$ in the pulse-to-pulse dataset. However, with the onset of
the dip $\sim 32$\,ks into the observation, the measured column density
increases drastically, first spiking to $\left( 24 \pm 2 \right) \times
10^{22}$\,cm$^{-2}$, quickly falling back to pre-dip levels, and then,
immediately before the peak of the flare, rising again to $\left( 16 \pm
2 \right) \times 10^{22}$\,cm$^{-2}$. We note that the partial-covering model in
the time-resolved spectrum of the dip reaches similar column densities ($\left(
19 \pm 3 \right) \times 10^{22}$\,cm$^{-2}$) to the pulse-to-pulse results,
suggesting that the necessity of the partial coverer may be due to the averaging
of spectra with varying \nh, rather than a spatial partial coverer. The spikes
in \nh suggest that the dip is caused by some overdense region of the accreted
stellar wind passing through the line of sight. The duration of the increases in
\nh along with the fact that the other continuum parameters do not change
significantly during the dip indicates that the material producing the
additional absorption must be relatively far from the star, as otherwise the
material would be accreted onto the star in a relatively short time, changing
the continuum emission and increasing the flux.

The continuum parameters (power-law index and cutoff and folding energies) are
generally flat for the first half of the observation, remaining this way until
a few pulsations before the peak of the flare, when the power-law index drops (in
both the time-resolved spectra and the pulse-to-pulse spectra) by a factor of
approximately 1/3. While the peak of the flare is clearly visible in the
lightcurve, the point where the power-law index drops is a better
approximation of the ``true'' starting point of the flare, with the beginning
stages of the flare being obscured by the second spike in \nh. The unabsorbed
flux is indeed climbing during the three pulses before the peak of the
flare, as can be seen in the overplotted lightcurve in Figure~\ref{fig:pulse}.
While the cutoff and folding energies do change slightly with the onset of the
flare, they are poorly constrained compared to the power-law index, so it is
difficult to see the effect of the flare in these parameters. The cutoff energy
does appear to drop during the flare and in the first post-flare spectrum,
before rising back to its pre-flare levels, while the even more poorly
constrained folding energy (and $kT$ in the \texttt{npex} model) rises during
the flare. However, these effects are statistically very weak.

The source after the flare is less luminous, in terms of unabsorbed flux, than
it was before the flare. In the pulse-to-pulse spectra, the power-law index
additionally shows an overall negative correlation with luminosity, with a slope
of $-0.067 \pm 0.008$ per $10^{-10}$\,erg\,cm$^{-2}$\,s$^{-1}$. Pearson's
coefficient for this is $-0.58$, which for 76 points corresponds to a two-tailed
p-value of less than $10^{-5}$.  This trend is preserved even if we exclude the
dip and the spike from the calculation, although the slope drops slightly to
$-0.058 \pm 0.011$ per $10^{-10}$\,erg\,cm$^{-2}$\,s$^{-1}$. The time-resolved
spectra show a similar correlation, with a slope of $-0.05 \pm 0.02$ per
$10^{-10}$\,erg\,cm$^{-2}$\,s$^{-1}$. This lends some credence to the
partial-covering model used in the final two spectra of the time-resolved
analysis --- without the partial-covering model, the power-law index in the
time-resolved spectra does not show a significant correlation with flux.

The continuum parameters generally show variability with phase, with the
different continuum models displaying similar behavior. The power-law index,
cutoff energy, folding energy, and \texttt{npex}'s $kT$ all show a peak
somewhere around the rising edge of the main pulse, and roughly trend downward
until the secondary pulse. The spectral shape of the rising and falling edges of
the main pulse are distinctly different. Our continuum results are similar to
those found using other satellites by, e.g., \citet{clark_discovery_1990},
\citet{coburn_magnetic_2001}, and \citet{robba_bepposax_2001}.

One potential area of concern is that the dip or flare represent some anomalous
behavior of the source, and including those data with the non-flaring,
less-obscured data may skew our results with regards any observed correlations.
However, the observed trends all persist even if we remove the dip and the flare
from the sample, indicating that the trends are somewhat fundamental to the
longer-term behavior of the source. The plot of iron line flux vs.\ power-law
flux with the dip and flare excluded has the same slope of $(1.1 \pm 0.2) \times
10^{-3}$ as found before, while the power-law index has a slightly smaller
slope, $-0.058 \pm 0.011$, but still consistent with the earlier result.
Removing the dip and flare obviously produces larger changes in the behavior of
\nh, but the overall appearance of the parameter when plotted against the
power-law flux is somewhat unchanged - there is an overall negative trend, as
one would expect, with increased scatter in the measured \nh as the luminosity
decreases. This increased scatter in \nh can be investigated by looking at the
(lack of a) correlation between \nh and the power-law index: periods of
increased \nh do not tend to be immediately related to changes in spectral
hardness.

\subsubsection{Connection between absorption event and flare}
\label{ssec:clump}

The presence of the flare in the lightcurve is evidence of some degree of
clumpiness or structure in the stellar wind of \qvnor \citep[although
simulations have shown that highly variable accretion can occur even in the
presence of a relatively smooth stellar wind; see,
e.g.,][]{manousakis_accretion_2014}. The increased \nh seen before the flare is
a possible source of the material accreted during the flare.  However, if this
is the case, the total amount of matter in the clump should reflect the observed
luminosity during the flare. The \nh variability has a two-peak structure, with
one large peak reaching $\sim 25 \times 10^{22}$\,cm$^{-2}$ several ks before
the flare and a smaller peak immediately before the flare that reaches $\sim 15
\times 10^{22}$\,cm$^{-2}$. By following a method similar to that applied to
Vela~X--1 by \citet{martinez_vela_2014}, we can estimate the size and, more
importantly, the mass of the overdensity in the stellar wind. We will first
assume that the spikes in \nh are caused by overdensities that move with the
stellar wind. Using the orbital solution from \citet{mukherjee_orbital_2006},
the \suz observation was carried out at an orbital phase of $\sim 0.3$, and so
we will additionally assume that the wind is moving roughly perpendicular to the
line of sight.

In general, the wind velocity at a given distance will depend on the
terminal wind velocity $v_{\infty}$ via
\begin{equation}
	v_{\rm wind}(r) = v_{\infty} \left( 1 - \frac{R_{\rm opt}}{r} \right)^{\beta}
	\label{eqn:vwind}
\end{equation}
The terminal wind velocity can in turn be estimated from the escape velocity for
\qvnor via the correlation plot in \citet{abbott_winds_1982}. From
\citet{clark_wind_1994}, we have
\begin{eqnarray}
	v_{\rm esc} &=& \left[2GM_{\rm opt}\left( 1-L_{\rm opt}/L_{\rm Edd} \right)/R_{\rm opt} \right]^{1/2} \\
							&\approx& 600\:\mathrm{km}\,\mathrm{s}^{-1}
	\label{eqn:vesc}
\end{eqnarray}
In these equations, $M_{\rm opt} \sim 15$\,\msol, $R_{\rm opt} \sim
16$\,\rsol, $\beta \sim 0.9$, $L_{\rm opt} = 1.6 \times 10^{5}$\,\lsol, and
$L_{\rm Edd} \sim 5 \times 10^{5}$\,\lsol are the mass, radius, Roche lobe
filling factor, luminosity, and Eddington luminosity of \qvnor
\citep{reynolds_optical_1538_1992,rawls_mass_2011}. From
\citet{abbott_winds_1982}, $v_{\infty} \sim 1300 - 1800$\,km\,s$^{-1}$.
A range of values have been reported for the mass and radius of \qvnor, with
\citet{reynolds_optical_1538_1992} finding $M_{\rm opt} = 19.9 \pm 3.4$\,\msol
and \citet{rawls_mass_2011} finding $20.72 \pm 2.27$\,\msol and $14.13 \pm
2.78$\,\msol for the elliptical and circular orbital solutions, respectively,
but these differences do not have large effects on our results here due to the
imprecision in estimating the terminal wind velocity.

At an inclination of $\sim 70$ degrees \citep{rawls_mass_2011}, the semimajor
axis of the binary system is $\sim 1.7 \times 10^{12}$\,cm, and at this distance
the wind velocity is $\sim 7 \times 10^{7}$\,cm\,s$^{-1}$. The length scales of
the overdensities can thus be estimated from their time of passage through the
line of sight from $v_{\rm wind}t_{\rm pass} \sim 2 \times 10^{11}$\,cm for the
large spike and $\sim 1 \times 10^{11}$\,cm for the smaller spike. The average
excess column densities over the extent of each spike in \nh are $9.1 \times
10^{22}$\,cm$^{-2}$ for the earlier, larger spike and $9.2 \times
10^{22}$\,cm$^{-2}$ for the second, smaller spike.  The similarity in average
column density is due to the shorter duration of the second spike ($\sim 3$
pulse periods) compared to the first spike ($\sim 7$ pulse periods). Given the
estimated length scales, these imply number densities of $\sim 4 \times
10^{11}$\,H atoms\,cm$^{-3}$ and $\sim 8 \times 10^{11}$\,H atoms\,cm$^{-3}$,
respectively, and thus masses of $\sim 10^{22}$\,g and $2 \times 10^{21}$\,g.
These represent a factor of $\sim$ 50--100 overdensity compared to the $\sim
5 \times 10^{9}$\,cm$^{-3}$ in the wind at the neutron star's distance from
\qvnor \citep{clark_wind_1994}.

The X-ray flare persists for $\sim 7$ pulse periods, if we define the ``flare''
state as when the source is spectrally harder (the green stars in Figure
\ref{fig:pulse}). If the material from the larger spike in \nh was fully
accreted by the neutron star, the mass accretion rate would be $\sim
10^{18}$\,g\,s$^{-1}$. If $\eta$, the efficiency of converting gravitational
energy to X-rays, is $\sim 10$\%, the X-ray luminosity would be
\begin{eqnarray}
	L_{\rm flare} &=& \eta\frac{GM_{\rm X}\dot{M}}{R_{\rm X}} \\
								&\approx& 3 \times 10^{37}\:\mathrm{erg}\,\mathrm{s}^{-1}
	\label{eqn:lflare}
\end{eqnarray}
The 5--100\,keV X-ray luminosity measured during the flare is somewhat lower
than this, with a peak luminosity of $1 \times 10^{37}$\,erg\,s$^{-1}$. However,
the structure of the \nh variability, with the second peak during the flare and
increased \nh post-flare, indicates that it is unlikely that the entirety of the
overdense region was accreted onto the neutron star during the flare.
Nonetheless, it is clear that the increased \nh seen during the dip represents
enough material to support the observed flaring.

\subsection{Iron line}
\label{ssec:iron}

In both the time-resolved and the pulse-to-pulse spectra, the Fe K$\alpha$ line
has a significant positive correlation with flux. The points from the
dip are slight outliers, remaining at roughly the same level as before the dip
during the first spike in \nh, and rising with the flare during the second,
smaller peak in \nh. The iron line's insensitivity to the dip is not surprising,
considering that the flux above $\sim 5$\,keV is largely unaffected by the large
increase in \nh that produces the dip. In the pulse-to-pulse spectra, Pearson's
correlation coefficient is 0.701, which indicates a very strong correlation.
A linear fit to the pulse-to-pulse data returns a slope of $\left( 1.1 \pm 0.2
\right)\times 10^{-2}$. The strong correlation between power-law flux and iron
line flux implies that the iron must be fairly close to the neutron star (i.e.
with a relatively small light travel time between the neutron star's continuum
emission and the surrounding iron). The cross-correlation of the pulse-to-pulse
iron line fluxes and power-law fluxes peaks at a lag of zero, which at our time
resolution places an upper limit on this light travel time of $525.59$\,s, or
a distance of $\sim 1.5 \times 10^{13}$\,cm. Meanwhile, phase-resolved
spectroscopy shows the iron line varying with phase, following the general shape
of the pulse profile, but with a significant phase lag of $\sim 0.67$, or $\sim
350$\,s. If we assume that the iron emission peaks due to the light from the
peak of the pulse, this would imply a distance of $\sim 10^{13}$\,cm, or $\sim
1$\,AU. Work by \citet{rodes-roca_detecting_2010} using \textit{XMM-Newton}
found variability of the iron line with phase, but they only report the iron
line equivalent width, and with the complex continuum that they use, a direct
comparison is difficult. \citet{robba_bepposax_2001} divided the
\textit{BeppoSAX} spectrum of \fif into four phase bins, and while they could
not constrain the intensity of the iron line especially well, their results are
broadly in line with ours, with the brightest iron K$\alpha$ offset from the
main pulse.

Our constraints on the iron K$\beta$ line provide some limited additional
insight in this area, as the ratio of the K$\beta$ intensity to the K$\alpha$
intensity is a probe of the level of ionization of the iron. While we typically
only find upper limits on the K$\beta$ intensity, the values we do find are
consistent with neutral iron, with the K$\beta$ to K$\alpha$ intensity ratio
sitting around the expected $13.5$\% \citep{palmeri_fe_2006}. This is again
consistent with the distance found via the cross-correlation and the
phase-resolved spectroscopy - if the iron is indeed at an average distance of
$\sim 1$\,AU, we should not expect to see any strong signature of ionization.

\subsection{Cyclotron scattering feature}
\label{ssec:crsf}
The behavior of the cyclotron line over the course of the observation is our
final feature of interest. This pseudo-absorption feature is produced when
photons are scattered out of the line of sight by electrons in the accretion
column above the neutron star's magnetic poles. In the $\sim 10^{12}$\,G
magnetic field of the neutron star, the cyclotron motion of the electrons is
quantized into Landau levels, and thus photons see an elevated scattering
cross-section when they have sufficient energy to send an electron into the next
level. This scattering results in a net reduction in line-of-sight photons at
around the CRSF energy. The ``12-B-12'' rule gives an approximate relation
between the observed CRSF energy and the magnetic field strength in the
scattering region:
\begin{equation}
	E_{\mathrm{cyc}} = \frac{11.57\,\mathrm{keV}}{1+z} \times B_{12}
	\label{eqn:12b12}
\end{equation}
where $B_{12}$ is the magnetic field in units of $10^{12}$\,G and $z \sim 0.15$
is the gravitational redshift in the scattering region. \fif's CRSF energy of
$\sim 23$\,keV implies a magnetic field strength of $\sim 2.2-2.3 \times
10^{12}$\,G, assuming a radius of $10$\,km and a mass of either $0.87 \pm
0.07$\,\msol or $1.104 \pm 0.177$\,\msol, per \citet{rawls_mass_2011}. Our
measured cyclotron line energy is moderately higher than many previously
measured values, with \citet{clark_discovery_1990}, \citet{robba_bepposax_2001},
and \citet{coburn_magnetic_2001} all finding CRSF energies around $\sim
20-21$\,keV, while our CRSF energy is closer to $\sim 23$\,keV. The results of
\citet{rodes-roca_first_2009} and \citet{hemphill_2013} are closer in line with
ours. While the possibility of a long-term change over time in the cyclotron
line energy is intriguing, a proper comparison with past work would need to take
much more careful account of the differing instruments and models (both for the
continuum and the CRSF) used by different authors, as these can have not
insignificant effects on the measured line energy \citep[see, e.g.
][]{muller_nocorrelation_2012}.

The cyclotron line energy reaches a peak value of $\sim 24$\,keV during the
flare, and there is a positive correlation between the CRSF energy and
luminosity. Our linear fits to the data find a slope of $0.3 \pm 0.2$\,keV per
$10^{-10}$\,erg\,cm$^{-2}$\,s$^{-1}$ in the \texttt{fdcut} and \texttt{highecut}
models and $0.4 \pm 0.2$\,keV per $10^{-10}$\,erg\,cm$^{-2}$\,s$^{-1}$ when
using \texttt{npex}. This is somewhat dependent on the inclusion of the point
from the flare: if the flare is excluded, the slope for the \texttt{fdcut}
continuum changes to $0.3 \pm 0.3$, which, while just barely consistent with
zero, is still notably inconsistent with a negative slope.

Current theory regarding the cyclotron line, according to
\citet{becker_spectral_2012}, finds that the CRSF energy should exhibit different
behavior with respect to source luminosity depending on the luminosity regime of
the source. When the accretion column is locally super-Eddington ($L
> L_{\mathrm{crit}}$ in \citeauthor{becker_spectral_2012}), increases in
luminosity should push the line-emitting region upward, towards lower magnetic
fields, while below this critical luminosity, the opposite should occur. At
still lower luminosities, the line-producing region should be relatively close
to the neutron star surface, and is not expected to change drastically with
changes in luminosity.

After computing the luminosity of \fif for each time-resolved spectrum from the
5--100\,keV flux assuming a distance of $6.4$\,kpc
\citep{reynolds_optical_1538_1992}, we can compare our results with the
predictions of \citet{becker_spectral_2012}. $L_{\rm crit}$ is given by equation
32\ in \citet{becker_spectral_2012}:
\begin{eqnarray}
	L_{\rm crit} &=& 1.49 \times 10^{37}{\rm erg\,s}^{-1} \left( \frac{\Lambda}{0.1} \right)^{-7/5} w^{-28/15} \nonumber \\
	&&\times \left( \frac{M}{1.4{\rm\,M}_{\odot}} \right)^{29/30} \left( \frac{R}{10{\rm\,km}} \right)^{1/10} \left( \frac{B_{\rm surf}}{10^{12}{\rm\,G}} \right)^{16/15}
	\label{eqn:lcrit}
\end{eqnarray}
Here, $M$, $R$, and $B$ are, respectively, the mass, radius, and surface
magnetic field strength of the neutron star, $w = 1$ characterizes the shape of the
photon spectrum inside the column (where there is assumed to be a mean photon
energy of $\bar{E} = wkT_{\rm eff}$), and $\Lambda$ characterizes the mode of
accretion. \citeauthor{becker_spectral_2012} only consider the case of $\Lambda
= 0.1$, which approximates disk accretion. With this value for $\Lambda$, \fif is
sub-critical, with the $L_{\rm coul}$ cut-off of \citet{becker_spectral_2012}
sitting approximately between the highest non-flaring luminosity and the flare,
and the \citet{becker_spectral_2012} predictions hold. However, \fif is believed
to be a wind-accreting source, rather than a disk accretor, and thus we should
choose $\Lambda = 1$ in our calculations. This presents us with an interesting
result: the value for $L_{\rm crit}$ drops considerably, in fact falling below
the $L_{\rm coul}$ cut-off. It is uncertain as to what the theory predicts in
this regime. This is the same as was found by \citet{fuerst_vela_2014} for Vela
X$-$1, another wind-accreting HMXB of similar luminosity and with a CRSF of
comparable energy ($\sim 25$\,keV).

The problematic parameter here is $\Lambda$, which is difficult to determine
precisely. The $w$ parameter is unlikely to diverge much from $w = 1$, as the
spectrum inside the accretion column is dominated by bremsstrahlung emission
\citep{becker_continuum_2007}, and our results do not change significantly when
$w$ is varied around its assumed value of 1. The $\Lambda$ parameter, on the
other hand, is tied to the \alfven radius $R_{\rm A}$, as $R_{\rm A}$ in the
disk of a disk-accreting source will be smaller than $R_{\rm A}$ for
a wind-accretor experiencing a similar $\dot{M}$.  \citet{fuerst_vela_2014}
suggest that a very narrow accretion column could cause a breakdown of the
theory in this case, and as the accretion column radius is tied to the \alfven
radius, differing values for $\Lambda$ could have an effect here. However, it is
also possible that the assumption of $\Lambda = 1$ is not precisely correct for
\fif; that is, the mode of accretion is not purely spherical. If the overdensity
that produces the flare possesses significant angular momentum relative to the
average angular momentum in the stellar wind, a transient disk may form during
the flare and change the value of $\Lambda$ from its value during the
non-flaring state of the source.

\begin{figure}[ht]
	\centering
	\plotone{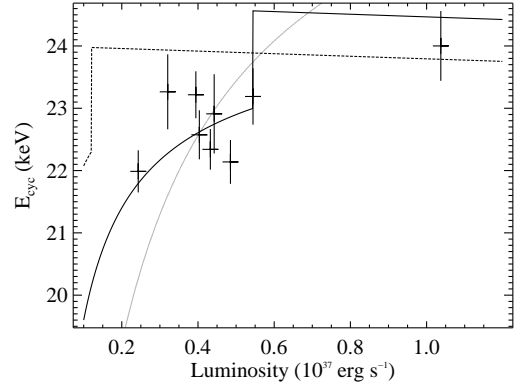}
	\caption{$E_{\rm cyc}$ versus luminosity for the \texttt{fdcut} continuum.
	Overplotted in a solid line is the fitted theoretical $E_{\rm cyc}$ from
	\citet{becker_spectral_2012}, where $\Lambda = 0.35$ and $E_{\rm surf}
	= 24.7$\,keV. Error bars are 1$\sigma$. The dotted line indicates the
	theoretical $E_{\rm cyc}$ assuming disk accretion ($\Lambda = 0.1$), while the
	dashed line is $E_{\rm cyc}$ assuming spherical wind accretion ($\Lambda
	= 1.0$). The sharp breaks in the curves for the fitted and wind-accreting
	cases are due to the switch from sub-critical to super-critical accretion at
	those luminosities.}
	\label{fig:ecyc_vs_theory}
\end{figure}

To examine this, we fit the predicted $E_{\rm cyc}$ from
\citet{becker_spectral_2012} to our cyclotron line energy measurements. The free
parameters were $\Lambda$ and the surface cyclotron line energy $E_{\rm surf}$,
although we limited $E_{\rm surf}$ to be greater than our largest measured
$E_{\rm cyc}$ (i.e., we assumed that the magnetic field our observation samples
is weaker than the magnetic field at the surface). The mass of the neutron star
was set to $1.104$\msol (although using the lower-mass estimate from
\citet{rawls_mass_2011} did not change our results significantly), and the
radius was assumed to be $10$\,km. As in \citet{becker_spectral_2012}, we assume
a dipole magnetic field. The theoretical $E_{\rm cyc}$ is thus dependent on the
altitude $h$ above the neutron star's surface:
\begin{equation}
	\frac{E_{\rm cyc}}{E_{\rm surf}} = \left( \frac{R + h}{R} \right)^{-3}
	\label{eqn:ecyc}
\end{equation}
where $h$ is given by 
\begin{eqnarray}
	h_{\rm s} &=& 2.28 \times 10^{3}{\rm\,cm} \left( \frac{\xi}{0.01} \right)\left( \frac{M}{1.4{\rm\,M}_{\odot}} \right)^{-1} \nonumber \\
	&& \times \left( \frac{R}{10{\rm\,km}} \right)\left( \frac{L_{\rm X}}{10^{37}{\rm\,erg\,s}^{-1}} \right)
		\label{eqn:hs}
\end{eqnarray}
for a supercritical source and by
\begin{eqnarray}
	h_{\rm c} &=&	1.48 \times 10^{5}{\rm\,cm} \left( \frac{\Lambda}{0.1} \right)^{-1} \left(\frac{\tau}{20} \right) \left( \frac{M}{1.4{\rm\,M}_{\odot}} \right)^{19/14} \nonumber \\
	&& \times \left(\frac{R}{10{\rm\,km}}\right)^{1/14} \left( \frac{B_{\rm surf}}{10^{12}{\rm\,G}}\right)^{-4/7} \left( \frac{L_{\rm X}}{10^{37}{\rm\,erg\,s}^{-1}}\right)^{-5/7}
	\label{eqn:hc}
\end{eqnarray}
for a subcritical source. This defines a piecewise function for $E_{\rm cyc}$,
which we fit to our results. For the \texttt{fdcut} continuum model, the fitted
values were $\Lambda = 0.348^{+0.002}_{-0.058}$ and $E_{\rm surf}
= 24.68^{+0.36}_{-0.03}$\,keV. This best-fit result is plotted over the measured
values for $E_{\rm cyc}$ in Figure~\ref{fig:ecyc_vs_theory}. We additionally fit
the predicted cyclotron line energy fixed $\Lambda = 1.0$ and $\Lambda = 0.1$.
While the quality of the $\Lambda = 0.348$ fit is poor ($\chi^{2} = 12.3$ for
7 degrees of freedom), it is clearly superior to the pure-disk and pure-wind
assumptions. The fitted function additionally implies a critical luminosity of
$\sim 6 \times 10^{36}$\,erg\,s$^{-1}$, in between the flaring and non-flaring
states of \fif.  This would suggest that during the flare, the luminosity
reached a sufficiently high level to create a stationary radiative shock in the
accretion column, which pushes down the average height of the line-producing
region. However, we lack data covering the transition from sub-critical to
super-critical accretion, and the discontinuous change in this formulation is
difficult to physically justify.

Alternative treatments of the CRSF-luminosity correlation can be
found in \citet{poutanen_reflection_2013} and \citet{nishimura_CRSF_2014}.
\citeauthor{poutanen_reflection_2013} posit that the CRSF is formed out of
light from the accretion column reflecting off of the neutron star surface,
finding that, for a dipole field, a higher-altitude emission region in the
accretion column will illuminate a larger fraction of the stellar surface and
sample a weaker average magnetic field. This results in smaller predicted
correlations compared to \citet{becker_spectral_2012}, but the signs of those
correlations are preserved, as the cyclotron line energy is still dependent on
the altitude of some scattering region in the accretion column. Their work
focuses on much higher luminosities than we see in \fif, even during the flare,
but their results should be somewhat compatible with our results, as their
observed trends are still dependent on variations in the altitude of the
emitting region in the accretion column. One area of concern is that the
\citeauthor{poutanen_reflection_2013} model suggests that the CRSF width should,
to some extent, vary inversely with the CRSF energy, as lower-energy lines are
produced by sampling a larger range of magnetic fields in the neutron star
atmosphere, smearing out the line. However, this interpretation is simplistic,
and more study is needed to determine exactly what the reflection model predicts
for the variation of line width with luminosity.

\begin{figure}[ht]
	\centering
	\plotone{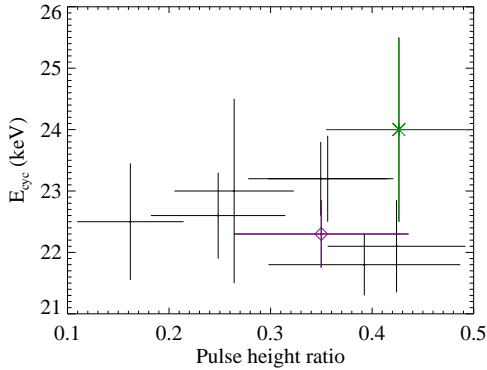}
	\caption{Cyclotron line energy vs.\ ratio of secondary to primary pulse height.
		Each two-peaked pulse of the source is fitted with two Gaussians, modeling
		the primary and secondary pulses, and the pulse height ratio is the ratio of
		the heights of the fitted Gaussians. The average pulse height ratio is then
		computed for each time bin used in Section \ref{ssec:time}. The dip and flare
		are indicated as a purple diamond and a green star, respectively.}
	\label{fig:ratio}
\end{figure}

Meanwhile, \citeauthor{nishimura_CRSF_2014} simulates
a superposition of cyclotron lines produced in the accretion column within
a range of altitudes, and manages to reproduce the observed correlations in
multiple sources. It is interesting to note here that, in the flare, the primary
pulse is considerably brighter than the secondary, and in our phase-resolved
analysis, the cyclotron line energy peaks with the primary pulse.  This may be
indicative of behavior similar to how \citeauthor{nishimura_CRSF_2014} explains
the observed positive correlation in Her~X$-$1: the increased primary pulse
relative to the rest of the pulse profile weights that phase bin's CRSF energy
above the others, and produces the observed positive trend. However, this would
imply that one should see a correlation between the ratio of the height of the
primary pulse to the height of the secondary pulse. By modeling each
double-peaked pulse with a sum of two Gaussians and computing the ratio of the
heights, we can confirm that this is not the case - the average pulse height
ratio for each time bin used in the time-resolved spectroscopy is not
significantly correlated with $E_{\mathrm{cyc}}$ (see Figure \ref{fig:ratio}). 

The CRSF width is also positively correlated with luminosity; this is in line
with the observed correlation between CRSF energy and width seen in multiple
sources and, indeed, across multiple sources \citep[see, e.g.,
][]{coburn_magnetic_2001}). The correlation between $E_{\rm cyc}$ and
$\sigma_{\rm cyc}$ is linear, with a slope of $0.9 \pm 0.6$. The ratio of
$\sigma_{\rm cyc}$ to $E_{\rm cyc}$ is, with the exception of the
lowest-luminosity measurement from time bin 8, constant at $\sim 0.11$. The
ratio of $E_{\rm cyc}$ to $\sigma_{\rm cyc}$ in a self-emitting atmosphere is
given by \citet{meszaros_comptonization_1985b} as

\begin{equation}
	\frac{\sigma_{\rm cyc}}{E_{\rm cyc}} \approx \left(8 \ln 2 \frac{kT_{\rm
	e}}{m_{\rm e}c^{2}}\right)^{\frac{1}{2}}|\cos\theta|
	\label{eqn:crsf_e_vs_w}
\end{equation}

As $\theta$ is the angle between the magnetic field and the line of sight,
$\cos\theta$ should not vary in phase-averaged spectra from a single
observation. Thus, the interpretation is that a constant $E_{\rm cyc}/\sigma_{\rm
cyc}$ ratio implies a relatively constant electron temperature of $\sim
1.3$\,keV. The largest outlier is time bin 8, where the width of the CRSF drops
to $\sim 1$\,keV. The line is easily detected in this spectrum, appearing
clearly in the residuals, but its shape is not fit well by the \texttt{gauabs}
model. The \texttt{cyclabs} model, which uses a Lorentzian line profile,
likewise does not fit the line well. This stems from the fact that there is no
currently available physics-based model for the CRSF, and so the fitted width in
these bins may not be physically meaningful. The CRSF depth is relatively stable
throughout the observation, within errors, although the low width in time bin
8 results in only a lower bound for the optical depth.

The cyclotron line varies significantly with phase (Figure \ref{fig:phase}), and
remains in-phase with the pulse profile in all continuum models. The behavior of
the cyclotron line energy and width are comparable to the results of
\citet{clark_discovery_1990} and \citet{coburn_magnetic_2001}. The peak value of
$E_{\mathrm{cyc}}$ is $\sim 24$\,keV, similar to the peak cyclotron line energy
seen in the time-resolved spectra. The CRSF width is also correlated with the
pulse profile, although this is a more conditional statement than it was for the
CRSF energy. The falling edge of the primary pulse is particularly problematic,
as the CRSF width in that phase bin is exceptionally low, at $\sim 0.1
- 0.3$\,keV. The depth in this bin is unbounded at the upper end due to the low
width, and the line is overall much more poorly detected in comparison to the
other phase bins, with its addition lowering the value of $\chi^{2}/{\rm dof}$
from $474.68/425$ to $445.04/423$ (compare to the peak of the main pulse, where
the CRSF lowers $\chi^{2}/{\rm dof}$ from $689.74/435$ to $495.89/432$).
However, beyond this one phase bin, the width shows a slight rise with phase,
peaking with the primary pulse, and there is a hint of a correlation between the
cyclotron line width and energy, similar to that seen in the phase-averaged
spectra, but weaker. The depth follows the opposite pattern, with the
highest-energy CRSF being one of the shallower features. The CRSF depth also
displays the same asymmetry across the main pulse as seen in the other spectral
parameters. Although Equation \ref{eqn:crsf_e_vs_w} predicts variation of
$\sigma_{\rm cyc}/E_{\rm cyc}$ with luminosity within the pulse, no significant
correlation can be found due to the large error bars.

\section{Summary \& Conclusions}
We have carried out the first \suz observation of \fif. Our comprehensive
spectral analysis of the source examines the time-dependent and
pulse-phase-dependent behavior of the source under a variety of spectral models.
The dip-and-flare structure midway through the observation is associated with
a significant increase in the line-of-sight absorption, with the timing and
ionization characteristics of the absorption increase supporting the
interpretation that the material that occulted the source was subsequently
accreted onto one of the magnetic poles of the source, producing the observed
flare. The intensity of the iron K$\alpha$ emission line at $6.4$\,keV is found
to correlate positively with luminosity in phase-averaged spectra, while the
power-law index displays a negative correlation.  A phase-resolved analysis
additionally finds significant phase dependence in the spectral parameters, most
notably finding a significant phase shift in the iron line intensity. The
luminosity and phase dependence of the iron line, along with the lack of
measurably high ionization, support the conclusion that the observed iron is, on
average, approximately $0.7$\,AU distant from the neutron star.

A positive correlation between the CRSF energy and luminosity is observed for
the first time in this source. This correlation is moderately well explained by
theoretical work by \citet{becker_spectral_2012}, with the source accreting at
sub-critical rates for most of the observation, increasing to above the critical
luminosity during the flare. This is the first work that examines the behavior
of the source at this high of a luminosity. Further work is necessary to
properly compare this work to previous observations of the source, due primarily
to differences in model choice and energy ranges used to calculate luminosity.

\acknowledgements
\label{sec:acknowlegements}
This research has made use of data and software obtained from NASA's High Energy
Astrophysics Science Archive Research Center (HEASARC), a service of Goddard
Space Flight Center and the Smithsonian Astrophysical Observatory. PBH was
supported by NASA grant NNX13AE68G.

\bibliographystyle{apj}
\bibliography{refs}

\end{document}